\documentclass[letterpaper]{IEEEtran}
\usepackage[T1]{fontenc}
\usepackage[latin9]{inputenc}
\usepackage{color}
\usepackage{amsmath}
\usepackage{amsthm}
\usepackage{amssymb}
\usepackage{graphicx}
\usepackage[unicode=true,
 bookmarks=true,bookmarksnumbered=true,bookmarksopen=true,bookmarksopenlevel=1,
 breaklinks=false,pdfborder={0 0 0},pdfborderstyle={},backref=false,colorlinks=true]
 {hyperref}
\hypersetup{pdftitle={Your Title},
 pdfauthor={Your Name},
 pdfpagelayout=OneColumn,pdfnewwindow=true,pdfstartview=XYZ,plainpages=false,linkcolor=blue,urlcolor=blue,citecolor=red,anchorcolor=blue}

\makeatletter

\pdfpageheight\paperheight
\pdfpagewidth\paperwidth

\theoremstyle{plain}
\newtheorem{thm}{\protect\theoremname}
\theoremstyle{definition}
\newtheorem{defn}[]{\protect\definitionname}
\theoremstyle{remark}
\newtheorem{rem}[]{\protect\remarkname}
\theoremstyle{plain}
\newtheorem{cor}[]{\protect\corollaryname}
\theoremstyle{plain}
\newtheorem{lem}[]{\protect\lemmaname}

\usepackage{bm}

\usepackage{subfigure}
\usepackage{epstopdf}
\usepackage{float}
\usepackage{subfloat}
\usepackage{array}
\usepackage{tabularx}
\usepackage{multirow}
\usepackage{tikz}
\usepackage{algorithmic}

\usepackage{authblk}

\tikzstyle{arw}=[->,>=latex]
\tikzstyle{node}=[draw,rectangle,rounded corners, minimum width=1cm,minimum height =.75 cm]

\usepackage{color}%May be necessary if you want to color links

\usepackage{adjustbox}

\usepackage{algorithmic}

\providecommand{\corollaryname}{Corollary}
\providecommand{\definitionname}{Definition}
\providecommand{\lemmaname}{Lemma}

\providecommand{\theoremname}{Theorem}
\newcounter{mytempeqncnt}

\providecommand{\corollaryname}{Corollary}
\providecommand{\definitionname}{Definition}
\providecommand{\lemmaname}{Lemma}
\providecommand{\remarkname}{Remark}
\providecommand{\theoremname}{Theorem}

\makeatother

\providecommand{\corollaryname}{Corollary}
\providecommand{\definitionname}{Definition}
\providecommand{\lemmaname}{Lemma}
\providecommand{\remarkname}{Remark}
\providecommand{\theoremname}{Theorem}

\begin{document}

\title{Distortion Bounds for Transmitting Correlated Sources with Common
Part over MAC }

\author{Lei Yu, Houqiang Li, \textit{Senior} \textit{Member, IEEE, }and Chang
Wen Chen, \textit{Fellow, IEEE} \thanks{Lei Yu and Houqiang Li are with the Department of Electronic Engineering
and Information Science, University of Science and Technology of China,
Hefei, China (e-mail: yulei@ustc.edu.cn, lihq@ustc.edu.cn). Chang
Wen Chen is with the Department of Computer Science and Engineering,
State University of New York at Buffalo, Buffalo, USA (e-mail: chencw@buffalo.edu). }}
\maketitle
\begin{abstract}
This paper investigates the joint source-channel coding problem of
sending two correlated memoryless sources with common part over a
memoryless multiple access channel (MAC). An inner bound and two outer
bounds on the achievable distortion region are derived. In particular,
they respectively recover the existing bounds for several special
cases, such as communication without common part, lossless communication,
and noiseless communication. When specialized to quadratic Gaussian
communication case, the inner bound and outer bound are used to generate
two new bounds. Numerical result shows that common part improves the
performance of such distributed communication system.
\end{abstract}

\section{Introduction}

The joint source-channel coding (JSCC) problem of transmitting correlated
sources (with common part) over multiple access channel was first
studied by Cover \emph{et al.} \cite{Cover} in which a bivariate
finite-alphabet source is to be transmitted losslessly over a two-to-one
multiple-access channel. As a lossy version of such JSCC problem,
Minero \emph{et al.} \cite{Minero} considered the achievable distortion
region of sending memoryless correlated source without common part
over memoryless multiple access channel, and they derived an inner
bound using a unified framework of hybrid coding (although the result
still holds for the communication with common part, however it will
become loose especially when the correlated sources are identical).
This unified hybrid coding \cite{Minero} generalizes the JSCC scheme
given by Cover \emph{et al.} \cite{Cover}, and can recover the achievability
result given in \cite{Cover}. In addition, specialized to quadratic
Gaussian communication case, the inner bound in \cite{Minero} can
also recover the performance of hybrid coding given by Lapidoth \emph{et
al.} \cite{Lapidoth}.

As for the converse part, Cover \emph{et al.} \cite{Cover} gave a
tight but uncomputable (multi-letter) outer bound for lossless communication
case, and Kang \emph{et al.} \cite{Kang} single-letterized this outer
bound by utilizing a data processing inequality on maximal correlation
coefficient. For lossy case, Lapidoth \emph{et al.} \cite{Lapidoth}
gave an outer bound for quadratic Gaussian communication utilizing
a similar data processing inequality as well. Recently, Lapidoth \emph{et
al.} \cite{Lapidoth16} also derived a new outer bound by the technology
of introducing an auxiliary random variable (or remote source). However,
the necessary condition of \cite{Lapidoth16} is weaker than the one
of \cite{Lapidoth} due to no data processing inequality applied in
the single-letterization processing \cite{Lapidoth16}.

As a lossy version of the problem (with common part) studied by Cover
\emph{et al.} \cite{Cover}, in this paper, we consider JSCC of transmitting
two memoryless correlated sources with common part over multiple access
channel, and give an inner bound and two outer bounds on the achievable
distortion region. For the inner bound, we propose an extended version
of hybrid coding by adding common part into the hybrid coding \cite{Minero}
that is designed for the case with no common part, and hence our inner
bound can recover the performance of the hybrid coding \cite{Minero}
by setting the common part to be empty. In addition, the outer bound
is derived by introducing auxiliary random variables (or remote sources)
as in \cite{Ozarow}-\cite{Khezeli}, and it can recover the existing
outer bounds when common part is absent at both encoders. When specialized
to Gaussian communication with Gaussian common part, our bounds reduce
to a new inner bound and a new outer bound.

The rest of this paper is organized as follows. Section II summarizes
basic notations, and formulates the problem. Section III gives the
main results for transmitting memoryless sources over memoryless MAC
problem. Section IV gives the main results for Gaussian communication
case. Finally, Section V gives the concluding remarks.

\section{Problem Formulation and Preliminaries}

\begin{figure}[t]
\centering\includegraphics[width=0.4\textwidth]{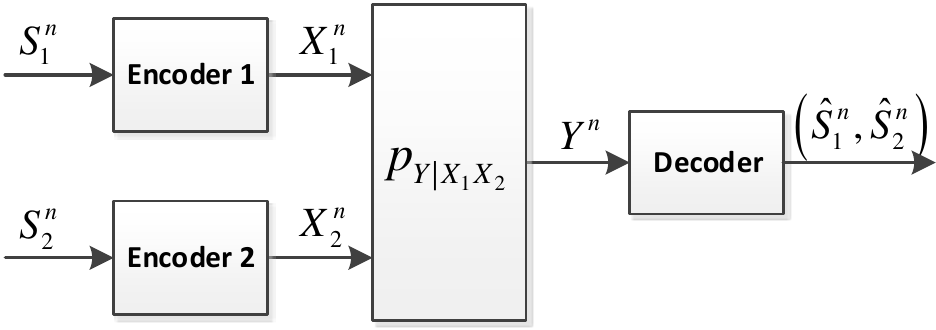} \caption{\label{fig:broadcast communication system }Communication of memoryless
correlated sources over a DM-MAC. }
\end{figure}

Consider the correlated sources $S_{1}$ and $S_{2}$ have common
part in sense of Gács-Körner-Witsenhausen common information \cite{G=00003D0000E1cs,Witsenhausen}.
\begin{defn}
$S_{0}$ is a common part of two correlated sources $S_{1}$ and $S_{2}$
if there exist two functions $f_{k}:\mathcal{S}_{k}\mapsto\mathcal{S}_{0},k=1,2$
such that $S_{0}=f_{1}\left(S_{1}\right)=f_{2}\left(S_{2}\right)$
with probability one, where $\mathcal{S}_{k}$ denotes the alphabet
of $S_{k}$, $k=0,1,2$. We say that $S_{1}$ and $S_{2}$ have a
common part if there exists a such $S_{0}$ as a common part of $S_{1}$
and $S_{2}$.
\end{defn}
Now consider the problem transmitting correlated sources over a multiple
access channel as shown in Fig. \ref{fig:broadcast communication system }.
The sender $k=1,2$ first codes discrete memoryless source $S_{k}^{n}$
into $X_{k}^{n}$ using a source-channel code, then transmits $X_{k}^{n}$
to a common receiver through a discrete memoryless multiple access
channel (DM-MAC) $p_{Y|X_{1},X_{2}}$, and finally, the receiver produces
source reconstructions $\hat{S}_{1}^{n}$ and $\hat{S}_{2}^{n}$ from
the received signal $Y^{n}$.

\begin{defn}
An $n$-length source-channel code is defined by the two encoding
functions $x_{k}^{n}:\mathcal{S}_{k}^{n}\mapsto\mathcal{X}_{k}^{n},k=1,2$
and two decoding functions $\hat{s}_{k}:\mathcal{Y}^{n}\mapsto\hat{\mathcal{S}_{k}^{n}},k=1,2$,
where $\hat{\mathcal{S}_{k}},\mathcal{X}_{k}$ and $\mathcal{Y}$
are the alphabet of source reconstruction $\hat{S_{k}}$, channel
input $X_{k}$, and channel output $Y$.
\end{defn}
For any $n$-length source-channel code, the induced distortion is
defined as
\begin{equation}
\mathbb{E}d_{k}(S_{k}^{n},\hat{S}_{k}^{n})=\frac{1}{n}\sum_{t=1}^{n}\mathbb{E}d_{k}(S_{k,t},\hat{S}_{k,t}),
\end{equation}
for $k=1,2$, where $d_{k}(s_{k},\hat{s}_{k}):{\mathcal{S}_{k}}\times\hat{{\mathcal{S}_{k}}}\mapsto\left[0,+\infty\right]$
is a distortion measure function for source $S_{k}$.
\begin{defn}
For transmitting sources $\left(S_{1},S_{2}\right)$ over MAC $p_{Y|X_{1},X_{2}}$,
we say the distortion tuple $\left(D_{1},D_{2}\right)$ is achievable,
if there exists a sequence of source-channel codes such that
\begin{equation}
\mathop{\limsup}\limits _{n\to\infty}\mathbb{E}d_{k}(S_{k}^{n},\hat{S}_{k}^{n})\le D_{k}.
\end{equation}
\end{defn}
\begin{defn}
For transmitting source $\left(S_{1},S_{2}\right)$ over MAC $p_{Y|X_{1},X_{2}}$,
the admissible distortion region is defined as
\begin{align}
\mathcal{R}\triangleq & \left\{ \left(D_{1},D_{2}\right):\left(D_{1},D_{2}\right)\textrm{ is achievable}\right\} .
\end{align}
In symmetric case,
\begin{align}
\mathcal{R}_{\textrm{sym}}\triangleq & \left\{ D:\left(D,D\right)\textrm{ is achievable}\right\} .
\end{align}
\end{defn}

\section{General Communication}

Now, we bound the distortion region for correlated sources communication
over MAC. We first define a distortion region {\small{}{}
\begin{align}
 & \mathcal{R}^{(i)}=\Bigl\{(D_{1},D_{2}):\textrm{There exist some pmf }p_{V_{0}|S_{0}}p_{V_{1}|S_{1},V_{0}}p_{V_{2}|S_{2},V_{0}},\nonumber \\
 & \text{and functions }x_{k}\left(v_{0},v_{k},s_{k}\right),\hat{s}_{k}\left(v_{0},v_{1},v_{2},y\right),k=1,2\text{ such that}\nonumber \\
 & \mathbb{E}d_{k}(S_{k},\hat{S}_{k})\le D_{k},k=1,2,\nonumber \\
 & I\left(V_{1};S_{1}|V_{0}V_{2}\right)<I\left(V_{1};Y|V_{0}V_{2}\right),\nonumber \\
 & I\left(V_{2};S_{2}|V_{0}V_{1}\right)<I\left(V_{2};Y|V_{0}V_{1}\right),\nonumber \\
 & I\left(V_{1}V_{2};S_{1}S_{2}|V_{0}\right)<I\left(V_{1}V_{2};Y|V_{0}\right),\nonumber \\
 & I\left(V_{0}V_{1}V_{2};S_{1}S_{2}\right)<I\left(V_{0}V_{1}V_{2};Y\right)\Bigr\}.\label{eq:innerbound}
\end{align}
}and another two distortion regions{\footnotesize{}{}}\footnote{{\footnotesize{}{}The $L$ in $\mathcal{R}_{1}^{(o)}$ is an arbitrary
positive integer.}}{\footnotesize{}{}
\begin{align}
 & \mathcal{R}_{1}^{(o)}=\nonumber \\
 & \Bigl\{(D_{1},D_{2}):\textrm{For any }p_{U_{[1:L]}|S_{1},S_{2}},\textrm{ there exist some pmf }p_{\hat{S}_{1},\hat{S}_{2}|S_{1},S_{2},U}\textrm{ and }\nonumber \\
 & p_{Q}\prod p_{S_{1},S_{2}}(s_{1,i},s_{2,i})p_{U_{[1:L]}|S_{1},S_{2}}(u_{[1:L],i}|s_{1,i},s_{2,i})p_{X_{1}|S_{1}^{n},Q}p_{X_{2}|S_{2}^{n},Q}\nonumber \\
 & \text{ such that }\mathbb{E}d_{k}(S_{k},\hat{S}_{k})\le D_{k},k=1,2,\nonumber \\
 & I(S_{1}S_{2};\hat{S}_{1}\hat{S}_{2}|U_{\mathcal{A}})\leq I\left(X_{1}X_{2};Y|U_{\mathcal{A}}^{n}Q\right)\textrm{ for any }\mathcal{A}\subseteq\left[1:L\right]\Bigr\}.
\end{align}
}and {\small{}{}
\begin{align}
 & \mathcal{R}_{2}^{(o)}=\nonumber \\
 & \Bigl\{(D_{1},D_{2}):\textrm{For any }p_{U|S_{1},S_{2}}\text{ such that }S_{1}\rightarrow\left(S_{0},U\right)\rightarrow S_{2},\nonumber \\
 & \textrm{ there exist some pmf }p_{\hat{S}_{1},\hat{S}_{2}|S_{1},S_{2},U}\textrm{ and }\nonumber \\
 & p_{Q}\prod p_{S_{1},S_{2}}\left(s_{1,i},s_{2,i}\right)p_{U|S_{1},S_{2}}\left(u_{i}|s_{1,i},s_{2,i}\right)p_{X_{1}|S_{1}^{n},Q}p_{X_{2}|S_{2}^{n},Q}\nonumber \\
 & \text{ such that }\mathbb{E}d_{k}(S_{k},\hat{S}_{k})\le D_{k},k=1,2,\nonumber \\
 & I(S_{1}S_{2};\hat{S}_{1}\hat{S}_{2})\leq I\left(X_{1}X_{2};Y|Q\right),\nonumber \\
 & I(S_{1}S_{2};\hat{S}_{1}\hat{S}_{2}|S_{0})\leq I\left(X_{1}X_{2};Y|S_{0}^{n}Q\right),\nonumber \\
 & I(S_{1};\hat{S}_{1}\hat{S}_{2}|S_{2})\leq I\left(X_{1};Y|X_{2}S_{2}^{n}Q\right),\nonumber \\
 & I(S_{2};\hat{S}_{1}\hat{S}_{2}|S_{1})\leq I\left(X_{2};Y|X_{1}S_{1}^{n}Q\right),\nonumber \\
 & I(S_{1}S_{2};\hat{S}_{1}\hat{S}_{2}|S_{0}U)\leq I\left(X_{1}X_{2};Y|S_{0}^{n}U^{n}Q\right),\nonumber \\
 & I(S_{1};\hat{S}_{1}\hat{S}_{2}|S_{2}U)\leq I\left(X_{1};Y|X_{2}S_{2}^{n}U^{n}Q\right),\nonumber \\
 & I(S_{2};\hat{S}_{1}\hat{S}_{2}|S_{1}U)\leq I\left(X_{2};Y|X_{1}S_{1}^{n}U^{n}Q\right)\Bigr\}.
\end{align}
}Note that in $\mathcal{R}_{2}^{(o)}$, $p_{U|S_{1},S_{2}}$ such
that $S_{1}\rightarrow\left(S_{0},U\right)\rightarrow S_{2}$ always
exists, since $S_{1}\rightarrow\left(S_{0},S_{1}\right)\rightarrow S_{2}$
and $S_{1}\rightarrow\left(S_{0},S_{2}\right)\rightarrow S_{2}$.
In addition, it is easy to verify that for such $p_{U|S_{1},S_{2}}$,
the random variables $\left(S_{1}^{n},S_{2}^{n},S_{0}^{n},U^{n},X_{1},X_{2},Q\right)$
in $\mathcal{R}_{2}^{(o)}$ satisfy $X_{1}\rightarrow\left(S_{1}^{n},Q\right)\rightarrow\left(S_{0}^{n},U^{n},Q\right)\rightarrow\left(S_{2}^{n},Q\right)\rightarrow X_{2}$.
Now we give the following theorem.
\begin{thm}
\label{thm:AdmissibleRegion-MAC} For transmitting sources $\left(S_{1},S_{2}\right)$
with common part $S_{0}$ over MAC $p_{Y|X_{1},X_{2}}$,
\begin{equation}
\mathcal{R}^{(i)}\subseteq\mathcal{R}\subseteq\mathcal{R}_{1}^{(o)}\subseteq\mathcal{R}_{2}^{(o)}.
\end{equation}
\end{thm}
\begin{rem}
The inner bound in Theorem \ref{thm:AdmissibleRegion-MAC} can be
easily extended to Gaussian or any other well-behaved continuous-alphabet
source-channel pair by standard discretization method \cite[Thm. 3.3]{El Gamal},
and moreover for this case the outer bound still holds. Theorem \ref{thm:AdmissibleRegion-MAC}
can be also extended to the case of source-channel bandwidth mismatch,
where $m$ samples of memoryless correlated sources are transmitted
through $n$ uses of a DM-MAC. This can be accomplished by replacing
the source and channel symbols in Theorem \ref{thm:AdmissibleRegion-MAC}
by supersymbols of lengths $m$ and $n$, respectively. Besides, Theorem
\ref{thm:AdmissibleRegion-MAC} can be also extended to the problem
with channel input cost (by adding channel input constraint).
\end{rem}
\begin{IEEEproof}
The proof of $\mathcal{R}^{(i)}\subseteq\mathcal{R}\subseteq\mathcal{R}_{1}^{(o)}$
is given in Appendix \ref{sec:MAC}. Now we show that $\mathcal{R}_{1}^{(o)}\subseteq\mathcal{R}_{2}^{(o)}$.
Actually $\mathcal{R}_{2}^{(o)}$ is a straightforward consequence
of $\mathcal{R}_{1}^{(o)}$. Choose $L=4,U_{1}=S_{1},U_{2}=S_{2},U_{3}=S_{0},U_{4}=U$
in $\mathcal{R}_{1}^{(o)}$, where $U|S_{1},S_{2}$ follows $p_{U|S_{1},S_{2}}$.
Then setting $\mathcal{A}=\emptyset,\left\{ 1\right\} ,\left\{ 2\right\} ,\left\{ 3\right\} ,\left\{ 1,4\right\} ,\left\{ 2,4\right\} ,\left\{ 3,4\right\} $
in $\mathcal{R}_{1}^{(o)}$ respectively gives the inequalities in
$\mathcal{R}_{2}^{(o)}$. Hence $\mathcal{R}_{1}^{(o)}\subseteq\mathcal{R}_{2}^{(o)}$.
\end{IEEEproof}
\begin{figure}[t]
\centering\includegraphics[width=1\columnwidth]{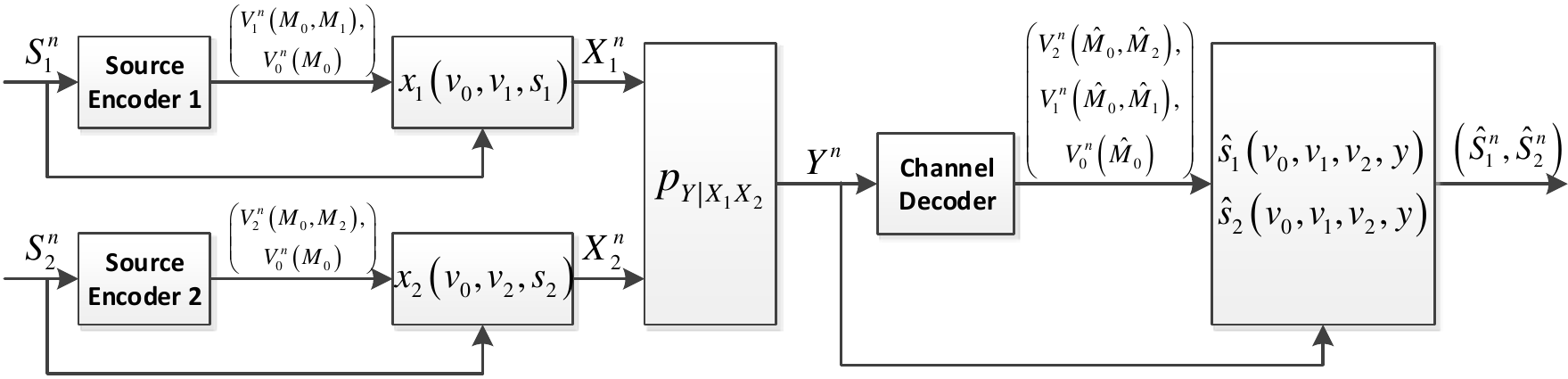} \caption{\label{fig:hybridcoding}The hybrid coding used to prove the inner
bound in Theorem \ref{thm:AdmissibleRegion-MAC}. }
%\vspace*{-0.1in}
\end{figure}

The inner bound $\mathcal{R}^{(i)}$ in Theorem \ref{thm:AdmissibleRegion-MAC}
is achieved by a unified hybrid coding scheme depicted in Fig. \ref{fig:hybridcoding}.
In this scheme, the codebook has a layered (or superposition) structure,
and consists of randomly and independently generated codewords $\left(V_{0}^{n}(m_{0}),V_{1}^{n}(m_{0},m_{1}),V_{2}^{n}(m_{0},m_{2})\right)$,
$(m_{0},m_{1},m_{2})\in\prod_{i=0}^{2}\left[1:2^{nr_{i}}\right]$,
where $(r_{0},r_{1},r_{2})$ denote the rates of the digital information
part and their values are given in Appendix \ref{sec:MAC}. At encoder
sides, upon source sequence $S_{k}^{n},k=1,2$, the encoder $k$ first
generates the common source $S_{0}^{n}$, and produces a common digital
message $M_{0}$ from $S_{0}^{n}$ by joint typicality encoding. Then
upon $M_{0}$ and $S_{k}^{n}$, the encoder $k$ produces a private
digital messages $M_{k}$. Finally, the codeword $\left(V_{0}^{n}(M_{0}),V_{k}^{n}(M_{0},M_{k})\right)$
and the source sequence $S_{k}^{n}$ are used to generate channel
input $X_{k}^{n}$ by symbol-by-symbol mapping $x_{k}\left(v_{0},v_{k},s_{k}\right)$.
At decoder side, upon received signal $Y^{n}$, the decoder reconstruct
$\left(M_{0},M_{1},M_{2}\right)$ (and also $\left(V_{0}^{n}(m_{0}),V_{1}^{n}(m_{0},m_{1}),V_{2}^{n}(m_{0},m_{2})\right)$)
losslessly by joint typicality decoding, and then it produces $\hat{S}_{k}^{n},k=1,2$
by symbol-by-symbol mapping $\hat{s}_{k}\left(v_{0},v_{1},v_{2},y\right)$.
Such a scheme could achieve any $(D_{1},D_{2})$ in the inner bound
$\mathcal{R}^{(i)}$.

Note that our hybrid coding is a extension of the hybrid coding in
\cite{Minero} to the case of correlated sources with common part.
By setting the common part as empty, i.e., $V_{0}=\emptyset$, our
result can recover the performance of the hybrid coding in \cite{Minero};
see the special cases in the following.

The outer bounds $\mathcal{R}_{1}^{(o)}$ and $\mathcal{R}_{2}^{(o)}$
in Theorem \ref{thm:AdmissibleRegion-MAC} are derived by introducing
auxiliary random variables (or remote sources) $U_{[1:L]}^{n}$ or
$U^{n}$. This bounding technology was originated in the multiple
description problem by Ozarow \cite{Ozarow}, and sequentially used
in the distributed source coding problem by Wagner \emph{et al.} \cite{Wagner-1}
and the source broadcast problem \cite{Reznic,Yu2016,Yu2016-1,Khezeli}.
As these works, in our proof one or multiple additional random variables
beyond those in the original problem are introduced. Besides, for
$\mathcal{R}_{2}^{(o)}$ the auxiliary random variable $U$ is restricted
to following the Markov chain $S_{1}\rightarrow\left(S_{0},U\right)\rightarrow S_{2}$.
This is inspired the work of Wagner \emph{et al.} \cite{Wagner-1},
where they showed that introducing an auxiliary random variable that
follows a Markov chain structure is sufficient to achieve the tight
bound for distributed source coding problem. Observe that our problem
is a extension of distributed source coding problem to the noisy communication
case, hence such a Markov chain structure is also used in our bound.
Besides, such a Markov chain structure also makes a sequence of data
processing inequalities available, which in turn generates some simpler
bounds (see Section \ref{sub:Outer-Bound}).

\subsection{Special Cases }
\begin{itemize}
\item Lossy Communication without Common Part
\end{itemize}
When there is no common part, the inner bound in Theorem \ref{thm:AdmissibleRegion-MAC}
reduces to \cite[Thm. 1]{Minero}, i.e.,{\small{}{}
\begin{align}
 & \mathcal{R}^{(i)}=\Bigl\{(D_{1},D_{2}):\textrm{There exist some pmf }p_{V_{0}}p_{V_{1}|S_{1},V_{0}}p_{V_{2}|S_{2},V_{0}},\nonumber \\
 & \text{and functions }x_{k}\left(v_{0},v_{k},s_{k}\right),\hat{s}_{k}\left(v_{0},v_{1},v_{2},y\right),k=1,2\text{ such that}\nonumber \\
 & \mathbb{E}d_{k}(S_{k},\hat{S}_{k})\le D_{k},k=1,2,\nonumber \\
 & I\left(V_{1};S_{1}|V_{2}V_{0}\right)<I\left(V_{1};Y|V_{2}V_{0}\right),\nonumber \\
 & I\left(V_{2};S_{2}|V_{1}V_{0}\right)<I\left(V_{2};Y|V_{1}V_{0}\right),\nonumber \\
 & I\left(V_{1}V_{2};S_{1}S_{2}|V_{0}\right)<I\left(V_{1}V_{2};Y|V_{0}\right)\Bigr\}.\label{eq:innerbound-6}
\end{align}
}In this case, $V_{0}$ is independent of $S_{1}$ and $S_{2}$, and
it becomes a timesharing auxiliary random variable. In addition, the
outer bound $\mathcal{R}_{2}^{(o)}$ reduces to a new outer bound{\small{}{}
\begin{align}
 & \mathcal{R}^{(o)}=\Bigl\{(D_{1},D_{2}):\textrm{For any }p_{U|S_{1},S_{2}}\text{ such that }S_{1}\rightarrow U\rightarrow S_{2},\nonumber \\
 & \textrm{there exist some pmf }p_{\hat{S}_{1},\hat{S}_{2}|S_{1},S_{2},U}\textrm{ and }\nonumber \\
 & p_{Q}\prod p_{S_{1},S_{2}}\left(s_{1,i},s_{2,i}\right)p_{U|S_{1},S_{2}}\left(u_{i}|s_{1,i},s_{2,i}\right)p_{X_{1}|S_{1}^{n},Q}p_{X_{2}|S_{2}^{n},Q}\nonumber \\
 & \text{such that }\mathbb{E}d_{k}(S_{k},\hat{S}_{k})\le D_{k},k=1,2,\nonumber \\
 & I(S_{1}S_{2};\hat{S}_{1}\hat{S}_{2})\leq I\left(X_{1}X_{2};Y|Q\right),\nonumber \\
 & I(S_{1};\hat{S}_{1}\hat{S}_{2}|S_{2})\leq I\left(X_{1};Y|X_{2}S_{2}^{n}Q\right),\nonumber \\
 & I(S_{2};\hat{S}_{1}\hat{S}_{2}|S_{1})\leq I\left(X_{2};Y|X_{1}S_{1}^{n}Q\right),\nonumber \\
 & I(S_{1}S_{2};\hat{S}_{1}\hat{S}_{2}|U)\leq I\left(X_{1}X_{2};Y|U^{n}Q\right),\nonumber \\
 & I(S_{1};\hat{S}_{1}\hat{S}_{2}|S_{2}U)\leq I\left(X_{1};Y|X_{2}S_{2}^{n}U^{n}Q\right),\nonumber \\
 & I(S_{2};\hat{S}_{1}\hat{S}_{2}|S_{1}U)\leq I\left(X_{2};Y|X_{1}S_{1}^{n}U^{n}Q\right)\Bigr\}.
\end{align}
}{\small \par}

\begin{itemize}
\item Lossless Communication with Common Part
\end{itemize}
When specialized to lossless communication of correlated source with
common part, by setting $V_{0}=\left(S_{0},W\right),V_{k}=\left(S_{k},X_{k}\right),x_{k}\left(v_{0},v_{k},s_{k}\right)=x_{k},\hat{s}_{k}\left(v_{0},v_{1},v_{2},y\right)=s_{k},k=1,2$
where $W$ is a random variable independent of $S_{1}$ and $S_{2}$,
the inner bound in Theorem \ref{thm:AdmissibleRegion-MAC} recovers
the inner bound \cite[Thm. 1]{Cover} on the admissible sources region,
i.e., {\small{}{}
\begin{align}
 & \mathcal{R}^{(i)}=\Bigl\{ p_{S_{1},S_{2}}:\textrm{There exist some pmf }p_{W}p_{X_{1}|S_{1},W}p_{X_{2}|S_{2},W}\text{ such that}\nonumber \\
 & H\left(S_{1}|S_{2}\right)\leq I\left(X_{1};Y|X_{2}S_{2}W\right),\nonumber \\
 & H\left(S_{2}|S_{1}\right)\leq I\left(X_{2};Y|X_{1}S_{1}W\right),\nonumber \\
 & H\left(S_{1}S_{2}|S_{0}\right)\leq I\left(X_{1}X_{2};Y|S_{0}W\right),\nonumber \\
 & H\left(S_{1}S_{2}\right)\leq I\left(X_{1}X_{2};Y\right)\Bigr\}.\label{eq:innerbound-1}
\end{align}
} In addition, the outer bound in Theorem \ref{thm:AdmissibleRegion-MAC}
reduces to a new outer bound {\small{}{}
\begin{align}
 & \mathcal{R}^{(o)}=\Bigl\{ p_{S_{1},S_{2}}:\textrm{For any }p_{U|S_{1},S_{2}}\text{ such that }S_{1}\rightarrow\left(S_{0},U\right)\rightarrow S_{2},\nonumber \\
 & \textrm{there exist some pmf }\nonumber \\
 & p_{Q}\prod p_{S_{1},S_{2}}\left(s_{1,i},s_{2,i}\right)p_{U|S_{1},S_{2}}\left(u_{i}|s_{1,i},s_{2,i}\right)p_{X_{1}|S_{1}^{n},Q}p_{X_{2}|S_{2}^{n},Q}\text{ such that}\nonumber \\
 & H\left(S_{1}S_{2}\right)\leq I\left(X_{1}X_{2};Y|Q\right),\nonumber \\
 & H\left(S_{1}|S_{2}\right)\leq I\left(X_{1};Y|X_{2}S_{2}^{n}Q\right),\nonumber \\
 & H\left(S_{2}|S_{1}\right)\leq I\left(X_{2};Y|X_{1}S_{1}^{n}Q\right),\nonumber \\
 & H\left(S_{1}S_{2}|S_{0}\right)\leq I\left(X_{1}X_{2};Y|S_{0}^{n}Q\right),\nonumber \\
 & H\left(S_{1}S_{2}|S_{0}U\right)\leq I\left(X_{1}X_{2};Y|S_{0}^{n}U^{n}Q\right),\nonumber \\
 & H\left(S_{1}|S_{2}U\right)\leq I\left(X_{1};Y|X_{2}S_{2}^{n}U^{n}Q\right),\nonumber \\
 & H\left(S_{2}|S_{1}U\right)\leq I\left(X_{2};Y|X_{1}S_{1}^{n}U^{n}Q\right)\Bigr\}.
\end{align}
}Furthermore, if $\left(S_{1},S_{2}\right)$ satisfy $S_{1}\rightarrow S_{0}\rightarrow S_{2}$,
then{\small{}{}
\begin{align}
\mathcal{R}=\mathcal{R}^{(i)}=\mathcal{R}^{(o)}= & \Bigl\{ p_{S_{1},S_{2}}:\textrm{There exist some pmf }\nonumber \\
 & p_{W}p_{X_{1}|S_{1},W}p_{X_{2}|S_{2},W}\text{ such that}\nonumber \\
 & H\left(S_{1}|S_{2}\right)\leq I\left(X_{1};Y|X_{2}S_{2}W\right),\nonumber \\
 & H\left(S_{2}|S_{1}\right)\leq I\left(X_{2};Y|X_{1}S_{1}W\right),\nonumber \\
 & H\left(S_{1}S_{2}|S_{0}\right)\leq I\left(X_{1}X_{2};Y|S_{0}W\right),\nonumber \\
 & H\left(S_{1}S_{2}\right)\leq I\left(X_{1}X_{2};Y\right)\Bigr\}.\label{eq:}
\end{align}
}This implies the admissible sources region for transmitting the correlated
sources that are conditionally independent given the common part has
been characterized completely. As a counterpart, the admissible sources
region for broadcasting conditionally independent sources has been
given in \cite{Khezeli}.

When $S_{0},S_{1},S_{2}$ are independent, the distortion region can
be used to derive the capacity region of Multiple Access Channel with
Common Message.
\begin{itemize}
\item Multiple Access Channel with Common Message
\end{itemize}
Consider lossless communication of independent sources $S_{0},S_{1},S_{2}$
with $H\left(S_{k}\right)=R_{k},k=0,1,2,$ then the problem becomes
Multiple Access Channel with Common Message \cite{Slepian}. Specialized
to this case, \eqref{eq:} reduces to the capacity region \cite[Eqn. 11]{Slepian},
i.e., {\small{}{}
\begin{align}
\mathcal{R}=\mathcal{R}^{(i)}=\mathcal{R}^{(o)}= & \Bigl\{(R_{0},R_{1},R_{2}):\textrm{There exist some pmf }\nonumber \\
 & p_{W}p_{X_{1}|W}p_{X_{2}|W}\text{ such that}\nonumber \\
 & R_{1}\leq I\left(X_{1};Y|X_{2}W\right),\nonumber \\
 & R_{2}\leq I\left(X_{2};Y|X_{1}W\right),\nonumber \\
 & R_{1}+R_{2}\leq I\left(X_{1}X_{2};Y|W\right),\nonumber \\
 & R_{0}+R_{1}+R_{2}\leq I\left(X_{1}X_{2};Y\right)\Bigr\}.
\end{align}
}{\small \par}

\begin{itemize}
\item Distributed Source Coding with Common Part
\end{itemize}
Consider the MAC $p_{Y|X_{1},X_{2}}$ is noiseless, i.e., $Y=\left(X_{1},X_{2}\right)$,
and constrain $H\left(X_{k}\right)\leq R_{k},k=1,2,$ then the problem
becomes Distributed Source Coding with Common Part \cite{Wagner}.
Set $V_{k}=\left(V_{k},X_{k}\right),x_{k}\left(v_{0},v_{k},s_{k}\right)=x_{k},\hat{s}_{k}\left(v_{0},v_{1},v_{2},y\right)=\hat{s}_{k}\left(v_{0},v_{1},v_{2}\right),k=1,2$,
where $X_{1}$ and $X_{2}$ are two random variables independent of
each other and other variables. Then the inner bound of Theorem \ref{thm:AdmissibleRegion-MAC}
recovers the inner bound \cite[Thm.1]{Wagner} on the achievable distortion
region, i.e., {\small{}{}
\begin{align}
\mathcal{R}^{(i)}= & \Bigl\{(D_{1},D_{2}):\textrm{There exist some pmf }p_{V_{0}|S_{0}}p_{V_{1}|S_{1},V_{0}}p_{V_{2}|S_{2},V_{0}},\nonumber \\
 & \text{and functions }\hat{s}_{k}\left(v_{0},v_{1},v_{2}\right),k=1,2\text{ such that}\nonumber \\
 & \mathbb{E}d_{k}(S_{k},\hat{S}_{k})\le D_{k},k=1,2,\nonumber \\
 & I\left(V_{1};S_{1}|V_{0}V_{2}\right)<R_{1},\nonumber \\
 & I\left(V_{2};S_{2}|V_{0}V_{1}\right)<R_{2},\nonumber \\
 & I\left(V_{0}V_{1}V_{2};S_{1}S_{2}\right)<R_{1}+R_{2}\Bigr\}.\label{eq:innerbound-7}
\end{align}
}{\small \par}

\section{Quadratic Gaussian Communication}

In this section, we apply the result for general communication to
the quadratic Gaussian communication case. Consider sending jointly
Gaussian sources $S_{k}=\left(S_{0},S_{k}^{\prime}\right),k=1,2$
with $\left(S_{0},S_{1}^{\prime},S_{2}^{\prime}\right)\sim\mathcal{N}\left(\mathbf{0},\Sigma_{\left(S_{0},S_{1}^{\prime},S_{2}^{\prime}\right)}\right)$
and\footnote{Throughout this paper, we use $\Sigma_{\left(X,Y\right)}$ to denote
the covariance of $\left(X,Y\right)$ and $\Sigma_{X,Y}$ to denote
the cross-covariance of $X$ and $Y$.}
\begin{equation}
\Sigma_{\left(S_{0},S_{1}^{\prime},S_{2}^{\prime}\right)}=\left(\begin{array}{ccc}
1 & \rho_{01} & \rho_{02}\\
\rho_{01} & 1 & \rho_{12}\\
\rho_{02} & \rho_{12} & 1
\end{array}\right)
\end{equation}
over a power-constrained Gaussian MAC $Y=X_{1}+X_{2}+Z$ with $\mathbb{E}\left(X_{k}^{2}\right)\leq P_{k},k=1,2$
and $Z\sim\mathcal{N}\left(0,1\right)$\footnote{For simplicity, we assume source variances are unit and so is the
channel noise power, which can cover general cases by scaling $P_{k}$
and $D_{k}$.}. We also assume distortion is measured by quadratic distortion function
on $S_{k}^{\prime},k=1,2$, i.e., $d_{k}(s_{k},\hat{s}_{k})=d(s_{k}^{\prime},\hat{s}_{k})\triangleq(s_{k}^{\prime}-\hat{s}_{k})^{2},k=1,2$,
and source bandwidth and channel bandwidth are matched.

Without loss of generality, $\left(S_{0},S_{1}^{\prime},S_{2}^{\prime}\right)$
can be expressed as
\begin{align}
S_{1}^{\prime} & =\rho_{01}S_{0}+\sqrt{1-\rho_{01}^{2}}U_{1},\\
S_{2}^{\prime} & =\rho_{02}S_{0}+\sqrt{1-\rho_{02}^{2}}U_{2},
\end{align}
with
\begin{align}
U_{1} & =\beta_{1}U+\sqrt{1-\beta_{1}^{2}}B_{1},\label{eq:-17}\\
U_{2} & =\beta_{2}U+\sqrt{1-\beta_{2}^{2}}B_{2}.\label{eq:-19}
\end{align}
where $U\sim\mathcal{N}\left(0,1\right)$ and $B_{k}\sim\mathcal{N}\left(0,1\right),k=1,2$
are mutually independent Gaussian variables and also independent of
$S_{0}$, and
\begin{align}
\beta_{1}\beta_{2} & =\frac{\rho_{12}-\rho_{01}\rho_{02}}{\sqrt{\left(1-\rho_{01}^{2}\right)\left(1-\rho_{02}^{2}\right)}}.
\end{align}
Obviously $S_{1}\rightarrow\left(S_{0},U\right)\rightarrow S_{2}$
holds.

\subsection{Hybrid Coding Scheme}

In the following, we obtain the performance of hybrid coding scheme
by specializing the inner bound of Theorem \ref{thm:AdmissibleRegion-MAC}.
Let
\begin{align}
V_{0} & =S_{0}+W_{0}\\
V_{k} & =F_{k}\left(S_{0},S_{k},V_{0}\right)^{T}+W_{k},k=1,2,
\end{align}
and set $x_{k}\left(v_{0},v_{k},s_{k}\right),k=1,2$ to the linear
functions
\begin{equation}
X_{k}=G_{k}\left(S_{0},S_{k},V_{0},V_{k}\right)^{T},k=1,2,
\end{equation}
where $W_{k}\sim\mathcal{N}\left(0,\omega_{k}\right),k=0,1,2$ are
mutually independent and also independent of $S_{k},k=0,1,2$, and
$F_{k}=\left(f_{k,1},f_{k,2},f_{k,3}\right)$ and $G_{k}=\left(g_{k,1},g_{k,2},g_{k,3},g_{k,4}\right)$
are two row vectors of coefficients.

This induces the relationship
\[
\left(S_{0},S_{1},S_{2},V_{0},V_{1},V_{2},Y\right)^{T}=A\left(S_{0},S_{1},S_{2},W_{0},W_{1},W_{2},Z\right)^{T},
\]
where $A$ is given in (25)
\begin{figure*}[!t]
\setcounter{mytempeqncnt}{\value{equation}} \setcounter{equation}{24}
\begin{equation}
A=\left(\begin{array}{ccccccc}
1 & 0 & 0 & 0 & 0 & 0 & 0\\
0 & 1 & 0 & 0 & 0 & 0 & 0\\
0 & 0 & 1 & 0 & 0 & 0 & 0\\
1 & 0 & 0 & 1 & 0 & 0 & 0\\
f_{1,1}+f_{1,3} & f_{1,2} & 0 & f_{1,3} & 1 & 0 & 0\\
f_{2,1}+f_{2,3} & 0 & f_{2,2} & f_{2,3} & 0 & 1 & 0\\
a_{71} & g_{1,2}+g_{1,4}f_{1,2} & g_{2,2}+g_{2,4}f_{2,2} & g_{1,3}+g_{1,4}f_{1,3}+g_{2,3}+g_{2,4}f_{2,3} & g_{1,4} & g_{2,4} & 1
\end{array}\right)\label{eqn_dbl_x}
\end{equation}
\setcounter{mytempeqncnt}{\value{equation}} \setcounter{equation}{\value{mytempeqncnt}}
\hrulefill{}\vspace*{4pt}

\end{figure*}
with
\[
a_{71}=g_{1,1}+g_{2,1}+g_{1,3}+g_{2,3}+g_{1,4}\left(f_{1,1}+f_{1,3}\right)+g_{2,4}\left(f_{2,1}+f_{2,3}\right).
\]
Hence the covariance of $\left(S_{0},S_{1},S_{2},V_{0},V_{1},V_{2},Y\right)$
is given by
\begin{equation}
\Sigma_{\left(S_{0},S_{1},S_{2},V_{0},V_{1},V_{2},Y\right)}=A\Sigma_{\left(S_{0},S_{1},S_{2},W_{0},W_{1},W_{2},Z\right)}A^{T}.
\end{equation}

Set $\hat{s}_{k}\left(v_{0},v_{1},v_{2},y\right),k=1,2$ to the linear
functions
\begin{equation}
\hat{S}_{k}=\Sigma_{S_{k},\left(V_{0},V_{1},V_{2},Y\right)}\Sigma_{\left(V_{0},V_{1},V_{2},Y\right)}^{-1}\left(V_{0},V_{1},V_{2},Y\right)^{T},
\end{equation}
then the covariance of error $E_{k}\triangleq S_{k}-\hat{S}_{k},k=1,2$
is given by
\[
\Sigma_{E_{k}}=\Sigma_{S_{k}}-\Sigma_{S_{k},\left(V_{0},V_{1},V_{2},Y\right)}\Sigma_{\left(V_{0},V_{1},V_{2},Y\right)}^{-1}\Sigma_{S_{k},\left(V_{0},V_{1},V_{2},Y\right)}^{T}.
\]
In addition, owing to power constraint,
\begin{equation}
\Sigma_{X_{k}}\leq P_{k},
\end{equation}
where
\begin{equation}
\Sigma_{X_{k}}=G_{k}\Sigma_{\left(S_{0},S_{k},V_{0},V_{k}\right)}G_{k}^{T}.
\end{equation}
Substitute these random variables and functions into $\mathcal{R}^{(i)}$
in Theorem \ref{thm:AdmissibleRegion-MAC}, then we get the performance
of the hybrid coding.
\begin{thm}
\label{thm:hybrid} For transmitting Gaussian source with common part
over Gaussian MAC,{\small{}{}
\begin{align}
\mathcal{R}\supseteq\mathcal{R}_{h}^{(i)}\triangleq & \Bigl\{(D_{1},D_{2}):\textrm{There exist }F_{k},G_{k},\omega_{0},\omega_{1},\omega_{2},k=1,2\nonumber \\
 & \text{such that }\Sigma_{E_{k}}\le D_{k},\Sigma_{X_{k}}\leq P_{k},k=1,2,\nonumber \\
 & \frac{\left|\Sigma_{\left(V_{0},V_{2},S_{0},S_{1}\right)}\right|}{\left|\Sigma_{\left(V_{0},V_{2},V_{1},S_{0},S_{1}\right)}\right|}<\frac{\left|\Sigma_{\left(V_{0},V_{2},Y\right)}\right|}{\left|\Sigma_{\left(V_{0},V_{2},V_{1},Y\right)}\right|},\nonumber \\
 & \frac{\left|\Sigma_{\left(V_{0},V_{1},S_{0},S_{2}\right)}\right|}{\left|\Sigma_{\left(V_{0},V_{2},V_{1},S_{0},S_{2}\right)}\right|}<\frac{\left|\Sigma_{\left(V_{0},V_{1},Y\right)}\right|}{\left|\Sigma_{\left(V_{0},V_{2},V_{1},Y\right)}\right|},\nonumber \\
 & \frac{\left|\Sigma_{\left(V_{0},S_{0},S_{1},S_{2}\right)}\right|}{\left|\Sigma_{\left(V_{0},V_{2},V_{1},S_{0},S_{1},S_{2}\right)}\right|}<\frac{\left|\Sigma_{\left(V_{0},Y\right)}\right|}{\left|\Sigma_{\left(V_{0},V_{2},V_{1},Y\right)}\right|},\nonumber \\
 & \frac{\left|\Sigma_{\left(S_{0},S_{1},S_{2}\right)}\right|}{\left|\Sigma_{\left(V_{0},V_{2},V_{1},S_{0},S_{1},S_{2}\right)}\right|}<\frac{\left|\Sigma_{Y}\right|}{\left|\Sigma_{\left(V_{0},V_{2},V_{1},Y\right)}\right|}\Bigr\}.\label{eq:innerbound-4}
\end{align}
}{\small \par}
\end{thm}
\begin{IEEEproof}
Substitute the random variables and functions set above into $\mathcal{R}^{(i)}$
of Theorem \ref{thm:AdmissibleRegion-MAC}, then
\[
I\left(V_{1};S_{1}|V_{0}V_{2}\right)=\frac{1}{2}\log\frac{\left|\Sigma_{\left(V_{0},V_{2},S_{1}\right)}\right|\left|\Sigma_{\left(V_{0},V_{2},V_{1}\right)}\right|}{\left|\Sigma_{\left(V_{0},V_{2},V_{1},S_{1}\right)}\right|\left|\Sigma_{\left(V_{0},V_{2}\right)}\right|}
\]
and
\[
I\left(V_{1};Y|V_{0}V_{2}\right)=\frac{1}{2}\log\frac{\left|\Sigma_{\left(V_{0},V_{2},Y\right)}\right|\left|\Sigma_{\left(V_{0},V_{2},V_{1}\right)}\right|}{\left|\Sigma_{\left(V_{0},V_{2},V_{1},Y\right)}\right|\left|\Sigma_{\left(V_{0},V_{2}\right)}\right|}.
\]
Hence the inequality $I\left(V_{1};S_{1}|V_{0}V_{2}\right)<I\left(V_{1};Y|V_{0}V_{2}\right)$
in $\mathcal{R}^{(i)}$ is equivalent to $\frac{\left|\Sigma_{\left(V_{0},V_{2},S_{0},S_{1}\right)}\right|}{\left|\Sigma_{\left(V_{0},V_{2},V_{1},S_{0},S_{1}\right)}\right|}<\frac{\left|\Sigma_{\left(V_{0},V_{2},Y\right)}\right|}{\left|\Sigma_{\left(V_{0},V_{2},V_{1},Y\right)}\right|}$.
Similarly, the last three inequalities in $\mathcal{R}^{(i)}$ are
equivalent to the last three inequalities in $\mathcal{R}_{h}^{(i)}$.
\end{IEEEproof}

\subsection{Uncoded Scheme}

Now we consider an uncoded scheme which adopts linear symbol-by-symbol
encoders
\begin{align}
X_{1} & =g_{10}S_{0}+g_{11}U_{1}\label{eq:-15}\\
X_{2} & =g_{20}S_{0}+g_{22}U_{2}\label{eq:-16}
\end{align}
and MMSE (minimum mean square error) decoders (which is optimal given
the encoder \eqref{eq:-15} and \eqref{eq:-16})
\begin{align*}
\hat{S}_{1} & =\mathbb{E}\left(S_{1}^{\prime}|Y\right)\\
 & =\frac{\left(\rho_{01}(g_{10}+g_{20})+g_{11}\sqrt{1-\rho_{01}^{2}}+g_{22}\frac{\rho_{12}-\rho_{01}\rho_{02}}{\sqrt{1-\rho_{02}^{2}}}\right)Y}{(g_{10}+g_{20})^{2}+g_{11}^{2}+g_{22}^{2}+2g_{11}g_{22}\frac{\rho_{12}-\rho_{01}\rho_{02}}{\sqrt{\left(1-\rho_{01}^{2}\right)\left(1-\rho_{02}^{2}\right)}}+1},\\
\hat{S}_{2} & =\mathbb{E}\left(S_{2}^{\prime}|Y\right)\\
 & =\frac{\left(\rho_{02}(g_{10}+g_{20})+g_{22}\sqrt{1-\rho_{02}^{2}}+g_{11}\frac{\rho_{12}-\rho_{01}\rho_{02}}{\sqrt{1-\rho_{01}^{2}}}\right)Y}{(g_{10}+g_{20})^{2}+g_{11}^{2}+g_{22}^{2}+2g_{11}g_{22}\frac{\rho_{12}-\rho_{01}\rho_{02}}{\sqrt{\left(1-\rho_{01}^{2}\right)\left(1-\rho_{02}^{2}\right)}}+1}
\end{align*}
where $\left(g_{k0},g_{kk}\right),k=1,2$ satisfy power constraint
$g_{k0}^{2}+g_{kk}^{2}\leq P_{k},k=1,2$. Note that such uncoded scheme
is a special case of hybrid coding above.
\begin{thm}
\label{thm:uncoded}The distortion pairs $\left(D_{1}^{u},D_{2}^{u}\right)$
resulting from the described uncoded scheme are given by{\small{}{}
\begin{align*}
D_{1}^{u} & =1-\frac{\left(\rho_{01}(g_{10}+g_{20})+g_{11}\sqrt{1-\rho_{01}^{2}}+g_{22}\frac{\rho_{12}-\rho_{01}\rho_{02}}{\sqrt{1-\rho_{02}^{2}}}\right)^{2}}{(g_{10}+g_{20})^{2}+g_{11}^{2}+g_{22}^{2}+2g_{11}g_{22}\frac{\rho_{12}-\rho_{01}\rho_{02}}{\sqrt{\left(1-\rho_{01}^{2}\right)\left(1-\rho_{02}^{2}\right)}}+1},\\
D_{2}^{u} & =1-\frac{\left(\rho_{02}(g_{10}+g_{20})+g_{22}\sqrt{1-\rho_{02}^{2}}+g_{11}\frac{\rho_{12}-\rho_{01}\rho_{02}}{\sqrt{1-\rho_{01}^{2}}}\right)^{2}}{(g_{10}+g_{20})^{2}+g_{11}^{2}+g_{22}^{2}+2g_{11}g_{22}\frac{\rho_{12}-\rho_{01}\rho_{02}}{\sqrt{\left(1-\rho_{01}^{2}\right)\left(1-\rho_{02}^{2}\right)}}+1}.
\end{align*}
}Hence
\begin{align*}
\mathcal{R}\supseteq\mathcal{R}_{u}^{(i)}\triangleq & \Bigl\{(D_{1},D_{2}):\textrm{There exist }\left(g_{k0},g_{kk}\right),k=1,2\text{ such that}\\
 & g_{k0}^{2}+g_{kk}^{2}\leq P_{k},k=1,2,D_{1}\geq D_{1}^{u},D_{2}\geq D_{2}^{u}\Bigr\}.
\end{align*}
\end{thm}

\subsection{Outer Bound\label{sub:Outer-Bound}}

Substitute the random variable $U$ such that \eqref{eq:-17} and
\eqref{eq:-19} into the outer bound $\mathcal{R}_{2}^{(o)}$ of Theorem
\ref{thm:AdmissibleRegion-MAC}, then the following outer bound on
Gaussian communication is recovered.
\begin{thm}
\label{thm:OuterBoundGG} For transmitting Gaussian source with common
part over Gaussian MAC,{\small{}{}
\begin{align}
 & \mathcal{R}\subseteq\mathcal{R}^{(o)}\triangleq\biggl\{(D_{1},D_{2}):\textrm{There exist some values }0\leq\hat{\rho}\leq1,\nonumber \\
 & 0\leq\hat{\rho}_{0}\leq\rho_{12|0}\triangleq\frac{\rho_{12}-\rho_{01}\rho_{02}}{\sqrt{\left(1-\rho_{01}^{2}\right)\left(1-\rho_{02}^{2}\right)}}\text{ such that for any }\rho_{12|0}\leq\beta_{1}\leq1,\nonumber \\
 & R_{S_{1}S_{2}}(D_{1},D_{2})\leq\frac{1}{2}\log\left(1+P_{1}+P_{2}+2\hat{\rho}\sqrt{P_{1}P_{2}}\right),\nonumber \\
 & R_{S_{1}S_{2}|S_{0}}(D_{1},D_{2})\nonumber \\
 & \qquad\leq\frac{1}{2}\log\biggl(1+\Bigl[\frac{1-\hat{\rho}^{2}}{1-\rho_{12|0}^{2}},1\Bigr]^{-}\left(P_{1}+P_{2}+2\hat{\rho}_{0}\sqrt{P_{1}P_{2}}\right)\biggr),\nonumber \\
 & \frac{\left(1-\rho_{01}^{2}\right)\left(1-\rho_{12|0}^{2}\right)}{D_{1}}\leq1+[1-\hat{\rho}^{2},1-\hat{\rho}_{0}^{2}]^{-}P_{1},\nonumber \\
 & \frac{\left(1-\rho_{02}^{2}\right)\left(1-\rho_{12|0}^{2}\right)}{D_{2}}\leq1+[1-\hat{\rho}^{2},1-\hat{\rho}_{0}^{2}]^{-}P_{2},\nonumber \\
 & \Bigl[\frac{\left(1-\rho_{01}^{2}\right)\left(1-\beta_{1}^{2}\right)}{D_{1}},1\Bigr]^{+}\Bigl[\frac{\left(1-\rho_{02}^{2}\right)\left(1-\beta_{2}^{2}\right)}{D_{2}},1\Bigr]^{+}\leq1+\nonumber \\
 & \qquad\Bigl[(1-\theta_{1}^{2})P_{1}+(1-\theta_{2}^{2})P_{2},\Bigl(1-\frac{\hat{\rho}_{0}^{2}}{\beta_{2}^{2}}\Bigr)P_{1}+\Bigl(1-\frac{\hat{\rho}_{0}^{2}}{\beta_{1}^{2}}\Bigr)P_{2}\Bigr]^{-},\nonumber \\
 & \frac{\left(1-\rho_{01}^{2}\right)\left(1-\beta_{1}^{2}\right)}{D_{1}}\leq1+\Bigl[1-\theta_{1}^{2},1-\frac{\hat{\rho}_{0}^{2}}{\beta_{2}^{2}}\Bigr]^{-}P_{1},\nonumber \\
 & \frac{\left(1-\rho_{02}^{2}\right)\left(1-\beta_{2}^{2}\right)}{D_{2}}\leq1+\Bigl[1-\theta_{2}^{2},1-\frac{\hat{\rho}_{0}^{2}}{\beta_{1}^{2}}\Bigr]^{-}P_{2},\nonumber \\
 & \text{ for some }\theta_{1},\theta_{2}\text{ such that }0\leq\theta_{1},\theta_{2}\leq1,\hat{\rho}\leq\theta_{1}\theta_{2}\biggr\},\label{eq:outerboundGaussian}
\end{align}
}where $[x,y]^{+}\triangleq\max\left\{ x,y\right\} ,$ $[x,y]^{-}\triangleq\min\left\{ x,y\right\} ,$
$\beta_{2}=\frac{\rho_{12}-\rho_{01}\rho_{02}}{\beta_{1}\sqrt{\left(1-\rho_{01}^{2}\right)\left(1-\rho_{02}^{2}\right)}},$
{\small{}{}
\begin{align*}
 & R_{S_{1}S_{2}}(D_{1},D_{2})\\
 & =\inf_{\begin{array}{c}
p_{\hat{S}_{1}\hat{S}_{2}|S_{1}S_{2}}:\mathbb{E}(S_{k}^{\prime}-\hat{S}_{k})^{2}\leq D_{k},k=1,2\end{array}}I(S_{1}S_{2};\hat{S}_{1}\hat{S}_{2})\\
 & =\begin{cases}
\frac{1}{2}\log^{+}\frac{1}{D_{1}},\qquad\textrm{if }\rho_{12}^{2}\geq\frac{1-D_{2}}{1-D_{1}};\\
\frac{1}{2}\log^{+}\frac{1-\rho_{12}^{2}}{D_{1}D_{2}},\qquad\textrm{if }\rho_{12}^{2}\leq\left(1-D_{1}\right)\left(1-D_{2}\right);\\
\frac{1}{2}\log^{+}\frac{1-\rho_{12}^{2}}{D_{1}D_{2}-\left(|\rho_{12}|-\sqrt{\left(1-D_{1}\right)\left(1-D_{2}\right)}\right)^{2}},\,\textrm{otherwise}
\end{cases}
\end{align*}
}with $\log^{+}x\triangleq\max\left\{ \log x,0\right\} ,$ under the
assumption that $D_{1}\leq D_{2}$, denotes the minimum sum rate needed
to achieve both $D_{1}$ and $D_{2}$ at the receiver when the encoders
cooperate to encode their observations \cite[Thm. III.1]{Lapidoth},
and{\small{}{}
\begin{align*}
 & R_{S_{1}S_{2}|S_{0}}(D_{1},D_{2})\\
 & =\inf_{\begin{array}{c}
p_{\hat{S}_{1}\hat{S}_{2}|S_{0}S_{1}S_{2}}:\mathbb{E}(S_{k}^{\prime}-\hat{S}_{k})^{2}\leq D_{k},k=1,2\end{array}}I(S_{1}S_{2};\hat{S}_{1}\hat{S}_{2}|S_{0})\\
 & =\inf_{\begin{array}{c}
p_{\hat{U}_{1}\hat{U}_{2}|U_{1}U_{2}}:\mathbb{E}(U_{k}-\hat{U}_{k})^{2}\leq D_{k},k=1,2\end{array}}I(U_{1}U_{2};\hat{U}_{1}\hat{U}_{2})\\
 & =\begin{cases}
\frac{1}{2}\log^{+}\frac{1}{D_{1}^{\prime}},\qquad\textrm{if }\rho_{12|0}^{2}\geq\frac{1-D_{2}^{\prime}}{1-D_{1}^{\prime}};\\
\frac{1}{2}\log^{+}\frac{1-\rho_{12|0}^{2}}{D_{1}^{\prime}D_{2}^{\prime}},\qquad\textrm{if }\rho_{12|0}^{2}\leq\left(1-D_{1}^{\prime}\right)\left(1-D_{2}^{\prime}\right);\\
\frac{1}{2}\log^{+}\frac{1-\rho_{12|0}^{2}}{D_{1}^{\prime}D_{2}^{\prime}-\left(|\rho_{12|0}|-\sqrt{\left(1-D_{1}^{\prime}\right)\left(1-D_{2}^{\prime}\right)}\right)^{2}},\,\textrm{otherwise}
\end{cases}
\end{align*}
}with
\begin{align*}
D_{1}^{\prime} & =\frac{D_{1}}{1-\rho_{01}^{2}},\\
D_{2}^{\prime} & =\frac{D_{2}}{1-\rho_{02}^{2}},
\end{align*}
under the assumption that $D_{1}^{\prime}\leq D_{2}^{\prime}$, denotes
the minimum sum rate needed to achieve both $D_{1}$ and $D_{2}$
at the receiver when the side information $S_{0}$ is available at
both the encoders and the decoder and the encoders cooperate to encode
their observations with help of $S_{0}$ .
\end{thm}
The proof of Theorem \ref{thm:OuterBoundGG} is given in Appendix
\ref{sec:broadcast-GaussianSI}. The Maximal Correlation Theory (Hirschfeld\textendash Gebelein\textendash Rényi
maximal correlation) is exploited in the proof. When $\rho_{01}=\rho_{02}=0$,
Theorem \ref{thm:OuterBoundGG} can recover the outer bound without
common part \cite[Thm. IV.1]{Lapidoth}. Besides, Theorem \ref{thm:OuterBoundGG}
can be extended to any other source-channel pair by following similar
steps to the proof.

\begin{figure}[t]
\centering\includegraphics[width=0.45\textwidth]{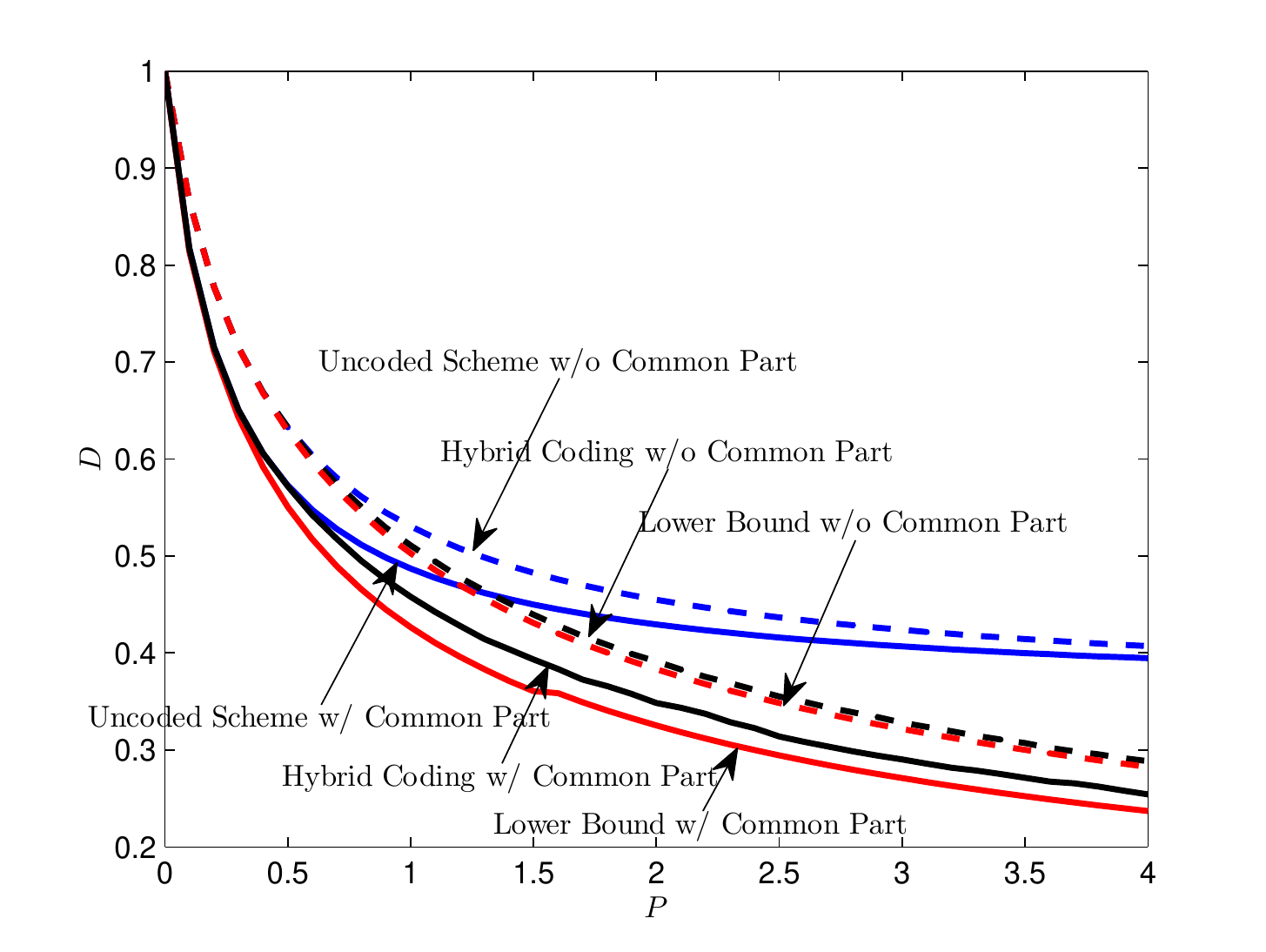} \protect\caption{\label{fig:HDAcoding-GG}Distortion bounds in the symmetric case for
sending Gaussian sources over Gaussian MAC. Uncoded Scheme w/ Common
Part, Hybrid Coding w/ Common Part, and Lower Bound w/ Common Part
are the bounds for the case with common part where $\rho_{12}=0.3,\rho_{01}=\rho_{02}=0.8$.
They respectively correspond to the inner bound in Theorem \ref{thm:uncoded},
the inner bound in Theorem \ref{thm:hybrid}, the outer bound in Corollary
\ref{thm:OuterBoundGG-1}. Uncoded Scheme w/o Common Part, Hybrid
Coding w/o Common Part, and Lower Bound w/o Common Part are the bounds
for the case with no common part where $\rho_{12}=0.3,\rho_{01}=\rho_{02}=0$.}
\end{figure}

When specialized to the symmetric case, Theorem \ref{thm:OuterBoundGG}
reduces to the following result.
\begin{cor}
\label{thm:OuterBoundGG-1} In the symmetric case, {\small{}{}
\begin{align}
 & \mathcal{R}_{\textrm{sym}}\subseteq\mathcal{R}_{\textrm{sym}}^{(o)}\triangleq\biggl\{ D:\textrm{There exist some values }0\leq\hat{\rho}\leq1,\nonumber \\
 & 0\leq\hat{\rho}_{0}\leq\rho_{12|0}\triangleq\frac{\rho_{12}-\rho_{01}\rho_{02}}{\sqrt{\left(1-\rho_{01}^{2}\right)\left(1-\rho_{02}^{2}\right)}}\text{ such that for any }\rho_{12|0}\leq\beta_{1}\leq1,\nonumber \\
 & R_{S_{1}S_{2}}(D,D)\leq\frac{1}{2}\log\left(1+2\left(1+\hat{\rho}\right)P\right),\nonumber \\
 & R_{S_{1}S_{2}|S_{0}}(D,D)\leq\frac{1}{2}\log\biggl(1+\Bigl[\frac{1-\hat{\rho}^{2}}{1-\rho_{12|0}^{2}},1\Bigr]^{-}\cdot2\left(1+\hat{\rho}_{0}\right)P\biggr),\nonumber \\
 & \frac{\left(1-\rho_{01}^{2}\right)\left(1-\rho_{12|0}^{2}\right)}{D}\leq1+[1-\hat{\rho}^{2},1-\hat{\rho}_{0}^{2}]^{-}P,\nonumber \\
 & \frac{\left(1-\rho_{02}^{2}\right)\left(1-\rho_{12|0}^{2}\right)}{D}\leq1+[1-\hat{\rho}^{2},1-\hat{\rho}_{0}^{2}]^{-}P,\nonumber \\
 & \Bigl[\frac{\left(1-\rho_{01}^{2}\right)\left(1-\beta_{1}^{2}\right)}{D},1\Bigr]^{+}\Bigl[\frac{\left(1-\rho_{02}^{2}\right)\left(1-\beta_{2}^{2}\right)}{D},1\Bigr]^{+}\nonumber \\
 & \qquad\leq1+\Bigl[2-\theta_{1}^{2}-\theta_{2}^{2},2-\frac{\hat{\rho}_{0}^{2}}{\beta_{2}^{2}}-\frac{\hat{\rho}_{0}^{2}}{\beta_{1}^{2}}\Bigr]^{-}P,\nonumber \\
 & \frac{\left(1-\rho_{01}^{2}\right)\left(1-\beta_{1}^{2}\right)}{D}\leq1+\Bigl[1-\theta_{1}^{2},1-\frac{\hat{\rho}_{0}^{2}}{\beta_{2}^{2}}\Bigr]^{-}P,\nonumber \\
 & \frac{\left(1-\rho_{02}^{2}\right)\left(1-\beta_{2}^{2}\right)}{D}\leq1+\Bigl[1-\theta_{2}^{2},1-\frac{\hat{\rho}_{0}^{2}}{\beta_{1}^{2}}\Bigr]^{-}P,\nonumber \\
 & \text{ for some }\theta_{1},\theta_{2}\text{ such that }0\leq\theta_{1},\theta_{2}\leq1,\hat{\rho}\leq\theta_{1}\theta_{2}\biggr\},\label{eq:outerboundGaussian-1}
\end{align}
}where
\begin{align*}
R_{S_{1}S_{2}}(D,D) & =\begin{cases}
\frac{1}{2}\log^{+}\frac{1-\rho_{12}^{2}}{D^{2}},\,\textrm{if }|\rho_{12}|\leq1-D;\\
\frac{1}{2}\log^{+}\frac{1+|\rho_{12}|}{2D-\left(1-|\rho_{12}|\right)},\,\textrm{otherwise}
\end{cases}
\end{align*}
and
\begin{align*}
R_{S_{1}S_{2}|S_{0}}(D,D) & =\begin{cases}
\frac{1}{2}\log^{+}\frac{1-\rho_{12|0}^{2}}{D^{\prime2}},\,\textrm{if }|\rho_{12|0}|\leq1-D^{\prime};\\
\frac{1}{2}\log^{+}\frac{1+|\rho_{12|0}|}{2D^{\prime}-\left(1-|\rho_{12|0}|\right)},\,\textrm{otherwise}
\end{cases}
\end{align*}
with
\begin{align*}
D^{\prime} & =\frac{D}{1-\rho_{01}^{2}}.
\end{align*}
\end{cor}
Fig. \ref{fig:HDAcoding-GG} illustrates the various bounds on the
achievable distortion.

\section{Concluding Remarks}

In this paper, we focused on the joint source-channel coding problem
of sending memoryless correlated sources with common part over memoryless
multiple access channel, and developed an inner bound and two outer
bounds for this problem. The inner bound is achieved by a unified
hybrid coding scheme with common part, and as special cases, it can
recover the performance of existing hybrid coding without common part.
Similarly, our outer bound can also recover several outer bounds in
the literature. When specialized to transmitting Gaussian sources
over Gaussian MAC, the inner bound and outer bound are used to generate
a new inner bound and a new outer bound, which can recover the best
known inner bound and outer bound without common part in the literature.

It is worth noting that in our results, two kinds of common informations
are involved. They are respectively in sense of Gács-Körner-Witsenhausen
common information \cite{G=00003D0000E1cs,Witsenhausen}, and in sense
of Wyner's common information \cite{Wyner}. In our problem, the
Gács-Körner-Witsenhausen common information, i.e., the common part,
has been exploited to improve the performance of communication system,
and the Wyner's common information has been exploited to obtain the
outer bounds. Besides, correlation ratio and maximal correlation coefficient
are also utilized to derive the outer bound for Gaussian communication
case. These concepts and tools are expected to be exploited to derive
achievability and converse results for other problems in network information
theory.

\appendices{}

\section{Proof of Theorem \ref{thm:AdmissibleRegion-MAC}\label{sec:MAC}}

\subsection{\label{sub:Inner-Bound}Inner Bound}

We use the hybrid coding shown in Fig. \ref{fig:hybridcoding} to
prove the inner bound.

\emph{Codebook Generation}: Fix conditional pmf $p_{V_{0}|S_{0}}p_{V_{1}|S_{1},V_{0}}p_{V_{2}|S_{2},V_{0}}$,
encoding functions $x_{k}\left(v_{0},v_{k},s_{k}\right),k=1,2$ and
decoding functions $\hat{s}_{k}\left(v_{0},v_{1},v_{2},y\right),k=1,2$
that satisfy all the inequalities in the inner bound \eqref{eq:innerbound}.
Randomly and independently generate a set of sequences $v_{0}^{n}(m_{0}),m_{0}\in[1:2^{nr_{0}}],$
with each distributed according to $\prod_{i=1}^{n}p_{V_{0}}(v_{0,i})$.
For $k=1,2$ and for each $m_{0}\in[1:2^{nr_{0}}]$, randomly and
independently generate a set of sequences $v_{k}^{n}(m_{0},m_{k}),m_{k}\in[1:2^{nr_{k}}],$
with each distributed according to $\prod_{i=1}^{n}p_{V_{k}|V_{0}}(v_{k,i}|v_{0,i}(m_{0}))$.
The codebook
\begin{align*}
\mathcal{C} & =\Bigl\{\left(v_{0}^{n}(m_{0}),v_{1}^{n}(m_{0},m_{1}),v_{2}^{n}(m_{0},m_{2})\right):\\
 & \qquad(m_{0},m_{1},m_{2})\in[1:2^{nr_{0}}]\times[1:2^{nr_{1}}]\times[1:2^{nr_{2}}]\Bigr\}.
\end{align*}
is revealed to both the encoders and the decoder.

\emph{Encoding}: We use joint typicality encoding. Let $\epsilon>\epsilon_{0}$.
Given $s_{0}^{n}$, both encoders 1 and 2 find the smallest index
$m_{0}$ such that $\left(s_{0}^{n},v_{0}^{n}\left(m_{0}\right)\right)\in\mathcal{T}_{\epsilon_{0}}^{\left(n\right)}$.
If there is no such index, let $m_{0}=1$. For $k=1,2$, given $s_{k}^{n}$
and $v_{0}^{n}\left(m_{0}\right)$, encoder $k$ finds the smallest
index $m_{k}$ such that $\left(s_{0}^{n},s_{k}^{n},v_{0}^{n}\left(m_{0}\right),v_{k}^{n}(m_{0},m_{k})\right)\in\mathcal{T}_{\epsilon}^{\left(n\right)}$.
If there is no such index, let $m_{k}=1$. Then the encoder $k$ transmits
the signal
\begin{equation}
x_{k,i}=x_{k}\left(v_{0,i}\left(m_{0}\right),v_{k,i}(m_{0},m_{k}),s_{k,i}\right),1\leq i\leq n.
\end{equation}

\emph{Decoding}: We use joint typicality decoding. Let $\epsilon'>\epsilon$.
Upon receiving signal $y^{n}$, the decoder of the receiver finds
the smallest index vector $(\hat{m}_{0},\hat{m}_{1},\hat{m}_{2})$
such that
\begin{equation}
(v_{0}^{n}(\hat{m}_{0}),v_{1}^{n}(\hat{m}_{0},\hat{m}_{1}),v_{2}^{n}(\hat{m}_{0},\hat{m}_{2}),y^{n})\in\mathcal{T}_{\epsilon'}^{\left(n\right)}.
\end{equation}
If there is no such index vector, let $(\hat{m}_{0},\hat{m}_{1},\hat{m}_{2})=(1,1,1)$.
The decoder reconstructs the sources as for $k=1,2$,
\begin{equation}
\hat{s}_{k,i}=\hat{s}_{k}(v_{0,i}(\hat{m}_{0}),v_{1,i}(\hat{m}_{0},\hat{m}_{1}),v_{2,i}(\hat{m}_{0},\hat{m}_{2}),y_{i}),1\leq i\leq n.
\end{equation}

\emph{Analysis of Expected Distortion}: We bound the distortion averaged
over $(S_{1}^{n},S_{2}^{n})$, and the random choice of the codebook
$\mathcal{C}$. Define the ``error\textquotedblright{} event {\small{}{}
\begin{align*}
 & \mathcal{E}=\\
 & \left\{ \left(S_{0}^{n},S_{1}^{n},S_{2}^{n},V_{0}^{n}(\hat{M}_{0}),V_{1}^{n}(\hat{M}_{0},\hat{M}_{1}),V_{2}^{n}(\hat{M}_{0},\hat{M}_{2}),Y^{n}\right)\notin\mathcal{T}_{\epsilon'}^{\left(n\right)}\right\} .
\end{align*}
}Then we have
\begin{align}
\mathcal{E} & \subseteq\mathcal{E}_{0}\cup\mathcal{E}_{1}\cup\mathcal{E}_{2}\cup\mathcal{E}_{3}\cup\mathcal{E}_{4}\cup\mathcal{E}_{5},
\end{align}
where{\small{}{}
\begin{align*}
\mathcal{E}_{0} & =\left\{ \left(S_{0}^{n},V_{0}^{n}(m_{0})\right)\notin\mathcal{T}_{\epsilon_{0}}^{\left(n\right)}\textrm{ for all }m_{0}\right\} ,\\
\mathcal{E}_{1} & =\left\{ \left(S_{0}^{n},S_{1}^{n},V_{0}^{n}(M_{0}),V_{1}^{n}(M_{0},m_{1})\right)\notin\mathcal{T}_{\epsilon}^{\left(n\right)}\textrm{ for all }m_{1}\right\} ,\\
\mathcal{E}_{2} & =\left\{ \left(S_{0}^{n},S_{2}^{n},V_{0}^{n}(M_{0}),V_{2}^{n}(M_{0},m_{2})\right)\notin\mathcal{T}_{\epsilon}^{\left(n\right)}\textrm{ for all }m_{2}\right\} ,\\
\mathcal{E}_{3} & =\left\{ \left(S_{0}^{n},S_{1}^{n},S_{2}^{n},V_{0}^{n}(M_{0}),V_{1}^{n}(M_{0},M_{1}),V_{2}^{n}(M_{0},M_{2})\right)\notin\mathcal{T}_{\epsilon_{1}}^{\left(n\right)}\right\} ,\\
\mathcal{E}_{4} & =\left\{ \left(S_{0}^{n},S_{1}^{n},S_{2}^{n},V_{0}^{n}(M_{0}),V_{1}^{n}(M_{0},M_{1}),V_{2}^{n}(M_{0},M_{2}),Y^{n}\right)\notin\mathcal{T}_{\epsilon'}^{\left(n\right)}\right\} ,\\
\mathcal{E}{}_{5} & =\Bigl\{\left(V_{0}^{n}(m'_{0}),V_{1}^{n}(m'_{0},m'_{1}),V_{2}^{n}(m'_{0},m'_{2}),Y^{n}\right)\in\mathcal{T}_{\epsilon'}^{\left(n\right)}\\
 & \qquad\qquad\textrm{ for some }(m'_{0},m'_{1},m'_{2})\neq(M_{0},M_{1},M_{2})\Bigr\},
\end{align*}
}for some $\epsilon_{1}$ such that $\epsilon_{0}<\epsilon<\epsilon_{1}<\epsilon'$.
Using union bound, we have
\begin{align}
\mathbb{P}\left(\mathcal{E}\right) & \leq\mathbb{P}\left(\mathcal{E}_{0}\right)+\mathbb{P}\left(\mathcal{E}_{0}^{c}\cap\mathcal{E}_{1}\right)+\mathbb{P}\left(\mathcal{E}_{0}^{c}\cap\mathcal{E}_{2}\right)\nonumber \\
 & +\mathbb{P}\left(\mathcal{E}_{0}^{c}\cap\mathcal{E}_{1}^{c}\cap\mathcal{E}_{2}^{c}\cap\mathcal{E}_{3}\right)+\mathbb{P}\left(\mathcal{E}_{3}^{c}\cap\mathcal{E}_{4}\right)+\mathbb{P}\left(\mathcal{E}_{5}\right).\label{eq:proberror}
\end{align}

Now we claim that if all the inequalities in the inner bound \eqref{eq:innerbound}
hold, then $\mathbb{P}\left(\mathcal{E}\right)$ tends to zero as
$n\to\infty$. Before proving it, we show that this claim implies
the distortions in the inner bound \eqref{eq:innerbound} are achievable.
The expected distortions are bounded by
\begin{align}
 & \limsup_{n\to\infty}\mathbb{E}d_{k}(S_{k}^{n},\hat{S}_{k}^{n})\nonumber \\
 & =\limsup_{n\to\infty}\Bigl(\mathbb{P}\left(\mathcal{E}_{4,k}\right)\mathbb{E}\left[d_{k}(S_{k}^{n},\hat{S}_{k}^{n})|\mathcal{E}_{4,k}\right]\nonumber \\
 & \qquad\qquad\qquad+\mathbb{P}\left(\mathcal{E}_{4,k}^{c}\right)\mathbb{E}\left[d_{k}(S_{k}^{n},\hat{S}_{k}^{n})|\mathcal{E}_{4,k}^{c}\right]\Bigr)\\
 & =\limsup_{n\to\infty}\mathbb{E}\left[d_{k}(S_{k}^{n},\hat{S}_{k}^{n})|\mathcal{E}_{4,k}^{c}\right]\\
 & \le\left(1+\epsilon'\right)\mathbb{E}d_{k}(S_{k},\hat{S}_{k})\label{eq:-1}\\
 & \le\left(1+\epsilon'\right)D_{k},
\end{align}
for $k=1,2$, where \eqref{eq:-1} follows from typical average lemma
\cite{El Gamal}. Therefore, the desired distortions are achieved
for sufficiently small $\epsilon'$.

Next we turn back to prove the claim above. Following from covering
lemma \cite[Sec. 3.7]{El Gamal}, the first three terms of \eqref{eq:proberror},
$\mathbb{P}\left(\mathcal{E}_{0}\right)+\mathbb{P}\left(\mathcal{E}_{0}^{c}\cap\mathcal{E}_{1}\right)+\mathbb{P}\left(\mathcal{E}_{0}^{c}\cap\mathcal{E}_{2}\right)$,
vanishes as $n\to\infty$ if
\begin{equation}
r_{0}>I\left(V_{0};S_{0}\right),\label{eq:-53}
\end{equation}
and according to Markov lemma \cite[Sec. 12.1.1]{El Gamal}, the fourth
item tends to zero as $n\to\infty$ if
\begin{equation}
r_{k}>I\left(V_{k};S_{k}|V_{0}\right),k=1,2.\label{eq:-62}
\end{equation}
Then by conditional typicality lemma \cite[Sec. 2.5]{El Gamal}, the
fifth item tends to zero as $n\to\infty$.

Now we focus on the last term of \eqref{eq:proberror}. $\mathcal{E}_{5}$
can be writen as
\begin{align}
\mathcal{E}_{5} & =\mathcal{E}_{51}\cup\mathcal{E}_{52}\cup\mathcal{E}_{53}\cup\mathcal{E}_{54},
\end{align}
where{\small{}{}
\begin{align*}
\mathcal{E}_{51} & =\Bigl\{\left(V_{0}^{n}(m'_{0}),V_{1}^{n}(m'_{0},m'_{1}),V_{2}^{n}(m'_{0},m'_{2}),Y^{n}\right)\in\mathcal{T}_{\epsilon'}^{\left(n\right)}\\
 & \qquad\qquad\textrm{ for some }m'_{0}\neq M_{0},m'_{1}\neq M_{1},m'_{2}\neq M_{2}\Bigr\},\\
\mathcal{E}_{52} & =\Bigl\{\left(V_{0}^{n}(M_{0}),V_{1}^{n}(M_{0},m'_{1}),V_{2}^{n}(M_{0},m'_{2}),Y^{n}\right)\in\mathcal{T}_{\epsilon'}^{\left(n\right)}\\
 & \qquad\qquad\textrm{ for some }m'_{1}\neq M_{1},m'_{2}\neq M_{2}\Bigr\},\\
\mathcal{E}_{53} & =\Bigl\{\left(V_{0}^{n}(M_{0}),V_{1}^{n}(M_{0},M_{1}),V_{2}^{n}(M_{0},m'_{2}),Y^{n}\right)\in\mathcal{T}_{\epsilon'}^{\left(n\right)}\\
 & \qquad\qquad\textrm{ for some }m'_{2}\neq M_{2}\Bigr\},\\
\mathcal{E}_{54} & =\Bigl\{\left(V_{0}^{n}(M_{0}),V_{1}^{n}(M_{0},m'_{1}),V_{2}^{n}(M_{0},M_{2}),Y^{n}\right)\in\mathcal{T}_{\epsilon'}^{\left(n\right)}\\
 & \qquad\qquad\textrm{ for some }m'_{1}\neq M_{1}\Bigr\},
\end{align*}
}Using union bound we have
\begin{align}
\mathbb{P}\left(\mathcal{E}_{5}\right) & \leq\mathbb{P}\left(\mathcal{E}_{51}\right)+\mathbb{P}\left(\mathcal{E}_{52}\right)+\mathbb{P}\left(\mathcal{E}_{53}\right)+\mathbb{P}\left(\mathcal{E}_{54}\right).\label{eq:proberror-1}
\end{align}
Following similar steps to the proof of \cite[Thm. 1]{Minero}, one
can prove $\mathbb{P}\left(\mathcal{E}_{51}\right)$ vanishes as $n\to\infty$
if
\begin{equation}
r_{0}+r_{1}+r_{2}<I\left(V_{0}V_{1}V_{2};Y\right)+I\left(V_{1};V_{2}|V_{0}\right),\label{eq:-63}
\end{equation}
$\mathbb{P}\left(\mathcal{E}_{52}\right)$ vanishes as $n\to\infty$
if
\begin{equation}
r_{1}+r_{2}<I\left(V_{1}V_{2};Y|V_{0}\right)+I\left(V_{1};V_{2}|V_{0}\right),\label{eq:-64}
\end{equation}
$\mathbb{P}\left(\mathcal{E}_{53}\right)$ vanishes as $n\to\infty$
if
\begin{equation}
r_{2}<I\left(V_{2};Y|V_{0}V_{1}\right),\label{eq:-65}
\end{equation}
and $\mathbb{P}\left(\mathcal{E}_{54}\right)$ vanishes as $n\to\infty$
if
\begin{equation}
r_{1}<I\left(V_{1};Y|V_{0}V_{2}\right).\label{eq:-66}
\end{equation}

Combining \eqref{eq:-53}, \eqref{eq:-62}, and \eqref{eq:-63}-\eqref{eq:-66}
leads to the sufficient condition, which completes the proof of the
inner bound.

\subsection{Outer Bound}

For fixed $p_{U_{[1:L]}|S_{1},S_{2}}$, we introduce a set of auxiliary
random variables $U_{[1:L]}^{n}$ that follow $\prod_{i=1}^{n}p_{U_{[1:L]}|S_{1},S_{2}}\left(u_{[1:L],i}|s_{1,i},s_{2,i}\right)$.
Then the Markov chain $U_{[1:L]}^{n}\rightarrow\left(S_{1}^{n},S_{2}^{n}\right)\rightarrow\left(X_{1}^{n},X_{2}^{n}\right)\rightarrow Y^{n}\rightarrow(\hat{S}_{1}^{n},\hat{S}_{2}^{n})$
holds. Assume $\mathcal{A}\subseteq\left[1:L\right]$. Next, we derive
a lower bound for $I\left(S_{1}^{n}S_{2}^{n};Y^{n}|U_{\mathcal{A}}^{n}\right)$.
{\small{}{}
\begin{align}
 & I\left(S_{1}^{n}S_{2}^{n};Y^{n}|U_{\mathcal{A}}^{n}\right)\nonumber \\
 & =\sum_{t=1}^{n}I\left(S_{1,t}S_{2,t};Y^{n}|U_{\mathcal{A}}^{n}S_{1}^{t-1}S_{2}^{t-1}\right)\\
 & =\sum_{t=1}^{n}H\left(S_{1,t}S_{2,t}|U_{\mathcal{A}}^{n}S_{1}^{t-1}S_{2}^{t-1}\right)\nonumber \\
 & \qquad\qquad-H\left(S_{1,t}S_{2,t}|Y^{n}U_{\mathcal{A}}^{n}S_{1}^{t-1}S_{2}^{t-1}\right)\label{eq:-14-3-1}\\
 & =\sum_{t=1}^{n}H\left(S_{1,t}S_{2,t}|U_{\mathcal{A},t}\right)-H\left(S_{1,t}S_{2,t}|Y^{n}U_{\mathcal{A}}^{n}S_{1}^{t-1}S_{2}^{t-1}\right)\label{eq:-3-1}\\
 & =\sum_{t=1}^{n}I\left(S_{1,t}S_{2,t};Y^{n}U_{\mathcal{A}}^{n}S_{1}^{t-1}S_{2}^{t-1}|U_{\mathcal{A},t}\right)\\
 & \geq\sum_{t=1}^{n}I(S_{1,t}S_{2,t};\hat{S}_{1,t}\hat{S}_{2,t}|U_{\mathcal{A},t})\\
 & =nI(S_{1,Q}S_{2,Q};\hat{S}_{1,Q}\hat{S}_{2,Q}|U_{\mathcal{A},Q}Q)\label{eq:-2-1}\\
 & =nI(S_{1,Q}S_{2,Q};\hat{S}_{1,Q}\hat{S}_{2,Q}Q|U_{\mathcal{A},Q})\\
 & \geq nI(S_{1,Q}S_{2,Q};\hat{S}_{1,Q}\hat{S}_{2,Q}|U_{\mathcal{A},Q})\\
 & =nI(S_{1}S_{2};\hat{S}_{1}\hat{S}_{2}|U_{\mathcal{A}}),\label{eq:-4-2}
\end{align}
}where $Q$ is a time-sharing random variable uniformly distributed
$\left[1:n\right]$ and independent of all other random variables,
and in \eqref{eq:-4-2}, $S_{k}\triangleq S_{k,Q},\hat{S}_{k}\triangleq\hat{S}_{k,Q},U_{l}\triangleq U_{l,Q},k=1,2,1\leq l\leq L$.

Now, we turn to upper-bounding $I\left(S_{1}^{n}S_{2}^{n};Y^{n}|U_{\mathcal{A}}^{n}\right)$.
\begin{align}
 & I\left(S_{1}^{n}S_{2}^{n};Y^{n}|U_{\mathcal{A}}^{n}\right)\nonumber \\
 & \leq I\left(X_{1}^{n}X_{2}^{n};Y^{n}|U_{\mathcal{A}}^{n}\right)\\
 & =\sum_{t=1}^{n}I\left(Y_{t};X_{1}^{n}X_{2}^{n}|U_{\mathcal{A}}^{n}Y^{t-1}\right)\\
 & \leq\sum_{t=1}^{n}I\left(Y_{t};X_{1}^{n}X_{2}^{n}Y^{t-1}|U_{\mathcal{A}}^{n}\right)\\
 & =\sum_{t=1}^{n}I\left(Y_{t};X_{1,t}X_{2,t}|U_{\mathcal{A}}^{n}\right)\label{eq:-5}\\
 & =nI\left(Y_{Q};X_{1,Q}X_{2,Q}|U_{\mathcal{A}}^{n}Q\right),\label{eq:-7}\\
 & =nI\left(Y;X_{1}X_{2}|U_{\mathcal{A}}^{n}Q\right),\label{eq:-14}
\end{align}
where \eqref{eq:-5} follows from $\left(X_{1}^{n},X_{2}^{n},Y^{t-1}\right)\rightarrow\left(X_{1,t},X_{2,t}\right)\rightarrow Y_{t}$,
$Q$ is the time-sharing random variable defined above, and $Y\triangleq Y_{Q},X_{k}\triangleq X_{k,Q},k=1,2$.

Combine \eqref{eq:-4-2} and \eqref{eq:-14}, then we have
\begin{equation}
I(S_{1}S_{2};\hat{S}_{1}\hat{S}_{2}|U_{\mathcal{A}})\leq I\left(X_{1}X_{2};Y|U_{\mathcal{A}}^{n}Q\right)\textrm{ for any }\mathcal{A}\subseteq\left[1:L\right].
\end{equation}

In addition, $\left(Q,S_{1}^{n},S_{2}^{n},X_{1},X_{2},Y\right)$ follows
the distribution $p_{Q}\prod p_{S_{1},S_{2}}\left(s_{1,i},s_{2,i}\right)p_{U_{[1:L]}|S_{1},S_{2}}\left(u_{[1:L],i}|s_{1,i},s_{2,i}\right)$
$p_{X_{1}|S_{1}^{n},Q}p_{X_{2}|S_{2}^{n},Q}p_{Y|X_{1},X_{2}}$. This
completes the proof of the outer bound $\mathcal{R}_{1}^{(o)}$.

\section{Proof of Theorem \ref{thm:OuterBoundGG}\label{sec:broadcast-GaussianSI}}

Before proving Theorem \ref{thm:OuterBoundGG}, we need introduce
several correlations and their properties, including correlation coefficient,
correlation ratio, maximal correlation coefficient, as well as the
corresponding conditional correlations.
\begin{defn}
For any random variables $W_{1}$ and $W_{2}$ with alphabets $\mathcal{W}_{1}\subseteq\mathbb{R}$
and $\mathcal{W}_{2}\subseteq\mathbb{R}$, the (Pearson) correlation
coefficient of $W_{1}$ and $W_{2}$ is defined by
\[
\rho(W_{1},W_{2})=\frac{\textrm{cov}\left(W_{1},W_{2}\right)}{\sqrt{\textrm{var}\left(W_{1}\right)}\sqrt{\textrm{var}\left(W_{2}\right)}}.
\]
Similarly, the conditional correlation coefficient of $W_{1}$ and
$W_{2}$ given another random variable $W_{0}$ is defined by
\begin{align*}
\rho(W_{1},W_{2}|W_{0}) & =\frac{\mathbb{E}\left[\textrm{cov}\left(W_{1},W_{2}|W_{0}\right)\right]}{\sqrt{\mathbb{E}\left[\textrm{var}\left(W_{1}|W_{0}\right)\right]}\sqrt{\mathbb{E}\left[\textrm{var}\left(W_{2}|W_{0}\right)\right]}}.
\end{align*}
\end{defn}
\begin{defn}
For any random variables $W_{1}$ and $W_{2}$ with alphabets $\mathcal{W}_{1}\subseteq\mathbb{R}$
and $\mathcal{W}_{2}$, the correlation ratio of $W_{1}$ on $W_{2}$
is defined by
\[
\theta\left(W_{1},W_{2}\right)=\sup_{f}\rho\left(W_{1},f(W_{2})\right),
\]
where the supremum is taken over all the functions $f:\mathcal{W}_{2}\mapsto\mathbb{R}$
satisfying
\begin{equation}
0<\mathbb{E}\left[f^{2}(W_{2})\right]<\infty.\label{eq:-2}
\end{equation}
Similarly, the conditional correlation ratio of $W_{1}$ on $W_{2}$
given another random variable $W_{0}$ is defined by
\[
\theta(W_{1},W_{2}|W_{0})=\sup_{f}\rho(W_{1},f(W_{2},W_{0})|W_{0}),
\]
where the supremum is taken over all the functions $f:\mathcal{W}_{2}\times\mathcal{W}_{0}\mapsto\mathbb{R}$
satisfying
\begin{equation}
0<\mathbb{E}\left[f^{2}(W_{2},W_{0})\right]<\infty.\label{eq:-2-2}
\end{equation}
\end{defn}
\begin{defn}
For any random variables $W_{1}$ and $W_{2}$ with alphabets $\mathcal{W}_{1}$
and $\mathcal{W}_{2}$, the maximal correlation coefficient of $W_{1}$
and $W_{2}$ is defined by
\[
\rho_{m}\left(W_{1},W_{2}\right)=\sup_{f_{1},f_{2}}\rho\left(f_{1}(W_{1}),f_{2}(W_{2})\right),
\]
where the supremum is taken over all the functions $f_{k}:\mathcal{W}_{k}\mapsto\mathbb{R}$
for $k=1,2$, satisfying
\begin{align}
0<\mathbb{E}\left[f_{k}^{2}(W_{k})\right] & <\infty,\label{eq:-3}
\end{align}
Moreover, the conditional maximal correlation coefficient of $W_{1}$
and $W_{2}$ given another random variable $W_{0}$ is defined by
\[
\rho_{m}(W_{1},W_{2}|W_{0})=\sup_{f_{1},f_{2}}\rho(f_{1}(W_{1},W_{0}),f_{2}(W_{2},W_{0})|W_{0}),
\]
where the supremum is taken over all the functions $f_{k}:\mathcal{W}_{k}\times\mathcal{W}_{0}\mapsto\mathbb{R}$
for $k=1,2$, satisfying
\begin{align}
0<\mathbb{E}\left[f_{k}^{2}(W_{k},W_{0})\right] & <\infty.\label{eq:-3-2}
\end{align}
\end{defn}
\begin{lem}
\cite{Yu16b}\label{lem:properties} For any random variables $W_{0}$,
$W_{1}$ and $W_{2}$, (conditional) correlation coefficient, (conditional)
correlation ratio, and (conditional) maximal correlation coefficient
have the following properties: {\small{}{}
\begin{equation}
0\leq\left|\rho\left(W_{1},W_{2}\right)\right|\leq\theta\left(W_{1},W_{2}\right)\leq\rho_{m}\left(W_{1},W_{2}\right)\leq1;
\end{equation}
\begin{align}
0\leq\left|\rho\left(W_{1},W_{2}|W_{0}\right)\right| & \leq\theta\left(W_{1},W_{2}|W_{0}\right)\nonumber \\
 & \qquad\qquad\leq\rho_{m}\left(W_{1},W_{2}|W_{0}\right)\leq1;
\end{align}
\begin{equation}
\theta\left(W_{1},W_{2}W_{0}\right)\geq\theta\left(W_{1},W_{0}\right);
\end{equation}
\begin{equation}
\rho_{m}\left(W_{1},W_{2}W_{0}\right)\geq\rho_{m}\left(W_{1},W_{0}\right);
\end{equation}
\begin{align}
\theta\left(W_{1},W_{2}\right) & =\sqrt{\frac{\textrm{var}\left(\mathbb{E}\left[W_{1}|W_{2}\right]\right)}{\textrm{var}\left(W_{1}\right)}}\nonumber \\
 & =\sqrt{1-\frac{\mathbb{E}\left[\textrm{var}\left(W_{1}|W_{2}\right)\right]}{\textrm{var}\left(W_{1}\right)}};\label{eq:e}
\end{align}
\begin{align}
\theta(W_{1},W_{2}|W_{0}) & =\sqrt{\frac{\mathbb{E}\left[\textrm{var}\left(\mathbb{E}\left[W_{1}|W_{2}W_{0}\right]|W_{0}\right)\right]}{\mathbb{E}\left[\textrm{var}\left(W_{1}|W_{0}\right)\right]}}\nonumber \\
 & =\sqrt{1-\frac{\mathbb{E}\left[\textrm{var}\left(W_{1}|W_{2}W_{0}\right)\right]}{\mathbb{E}\left[\textrm{var}\left(W_{1}|W_{0}\right)\right]}};
\end{align}
\begin{align}
\rho_{m}\left(W_{1},W_{2}\right) & =\sup_{f}\sqrt{\frac{\textrm{var}\left(\mathbb{E}\left[f(W_{1})|W_{2}\right]\right)}{\textrm{var}\left(f(W_{1})\right)}}\nonumber \\
 & =\sup_{f}\sqrt{1-\frac{\mathbb{E}\left[\textrm{var}\left(f(W_{1})|W_{2}\right)\right]}{\textrm{var}\left(f(W_{1})\right)}};
\end{align}
\begin{align}
\rho_{m}(W_{1},W_{2}|W_{0}) & =\sup_{f}\sqrt{\frac{\mathbb{E}\left[\textrm{var}\left(\mathbb{E}\left[f(W_{1},W_{0})|W_{2}W_{0}\right]|W_{0}\right)\right]}{\mathbb{E}\left[\textrm{var}\left(f(W_{1},W_{0})|W_{0}\right)\right]}}\nonumber \\
 & =\sup_{f}\sqrt{1-\frac{\mathbb{E}\left[\textrm{var}\left(f(W_{1},W_{0})|W_{2}W_{0}\right)\right]}{\mathbb{E}\left[\textrm{var}\left(f(W_{1},W_{0})|W_{0}\right)\right]}};
\end{align}
\begin{align}
 & 1-\theta^{2}\left(W_{1},W_{2}W_{0}\right)\nonumber \\
 & \qquad=\left(1-\theta^{2}\left(W_{1},W_{0}\right)\right)\left(1-\theta^{2}\left(W_{1},W_{2}|W_{0}\right)\right);\label{eq:i}
\end{align}
and
\begin{align}
 & 1-\theta^{2}\left(W_{1},W_{2}W_{0}|Z\right)\nonumber \\
 & \qquad=\left(1-\theta^{2}\left(W_{1},W_{0}|Z\right)\right)\left(1-\theta^{2}\left(W_{1},W_{2}|W_{0}Z\right)\right).
\end{align}
}
\end{lem}
Besides, some other remarkable properties are also needed in proving
Theorem \ref{thm:OuterBoundGG}.
\begin{lem}
\label{lem:iid}\cite[Thm. 1]{Witsenhausen} For a sequence of pairs
of independent random variables $\left(W_{1,i},W_{2,i}\right)_{i=1}^{n}$,
we have
\begin{equation}
\rho_{m}\left(W_{1}^{n},W_{2}^{n}\right)\leq\sup_{1\leq i\leq n}\rho_{m}\left(W_{1,i},W_{2,i}\right),\label{eq:-22}
\end{equation}
where $W_{k}^{n}=\left(W_{k,1},W_{k,2},\cdots,W_{k,n}\right)$ for
$k=1,2$.
\end{lem}
\begin{lem}
\label{lem:Gaussian}\cite[Sec. IV, Lem. 10.2]{Rozanov} For jointly
Gaussian random variables $W_{0},W_{1}$ and $W_{2}$, we have
\begin{align}
\rho_{m}\left(W_{1},W_{2}\right) & =|\rho\left(W_{1},W_{2}\right)|,\\
\rho_{m}\left(W_{1},W_{2}|W_{0}\right) & =|\rho\left(W_{1},W_{2}|W_{0}\right)|.
\end{align}
\end{lem}
\begin{lem}[Data Processing Inequality]
\cite{Yu16b}\label{lem:dpi} If random variable $W$ and non-degenerate
random variables $X,Y,Z$ \emph{form a Markov chain $X\rightarrow\left(Y,W\right)\rightarrow Z$,
then }
\begin{align}
\rho(X,Z|W) & \leq\theta\left(X,Y|W\right)\theta\left(Z,Y|W\right),\label{eq:-30}\\
\theta(X,Z|W) & \leq\theta\left(X,Y|W\right)\rho_{m}\left(Z,Y|W\right),\\
\rho_{m}(X,Z|W) & \leq\rho_{m}\left(X,Y|W\right)\rho_{m}\left(Z,Y|W\right).\label{eq:-32}
\end{align}
Moreover, the equalities hold in \eqref{eq:-30}-\eqref{eq:-32},
if $(X,Y,W)$ and $(Z,Y,W)$ have the same distribution. In particular,
if $W$ is degenerate, then
\begin{align}
\rho(X,Z) & \leq\theta\left(X,Y\right)\theta\left(Z,Y\right),\label{eq:-30-1}\\
\theta(X,Z) & \leq\theta\left(X,Y\right)\rho_{m}\left(Z,Y\right),\\
\rho_{m}(X,Z) & \leq\rho_{m}\left(X,Y\right)\rho_{m}\left(Z,Y\right).\label{eq:-32-1}
\end{align}
\end{lem}
Now we use $\mathcal{R}_{2}^{(o)}$ to prove Theorem \ref{thm:OuterBoundGG}.
For $\mathcal{R}_{2}^{(o)}$, denote $\hat{\rho}$ as the correlation
coefficient between $X_{1}$ and $X_{2}$, i.e., $\hat{\rho}\triangleq\rho(X_{1},X_{2})$,
and $\theta_{k}$ as correlation ratio of $X_{k}$ on $\left(S_{0}^{n},U^{n},Q\right)$
, i.e.,
\begin{equation}
\theta_{k}\triangleq\theta\left(X_{k},S_{0}^{n}U^{n}Q\right),k=1,2.\label{eq:-24}
\end{equation}
It should hold that $0\leq\hat{\rho},\theta_{1},\theta_{2}\leq1$.
Observe that in $\mathcal{R}_{2}^{(o)}$, $X_{1}\rightarrow\left(S_{1}^{n},Q\right)\rightarrow\left(S_{0}^{n},U^{n},Q\right)\rightarrow\left(S_{2}^{n},Q\right)\rightarrow X_{2}$
holds. Hence from Lemma \ref{lem:dpi}, we have
\begin{equation}
\hat{\rho}\leq\theta_{1}\theta_{2}.
\end{equation}
From Property \eqref{eq:e} of Lemma \ref{lem:properties}, we have
\begin{align}
\mathbb{E}\left[\textrm{var}(X_{k}|S_{0}^{n}U^{n}Q)\right] & =\left(1-\theta_{k}^{2}\right)\mathbb{E}\left[\textrm{var}(X_{k})\right],k=1,2.
\end{align}

In addition, denote $\hat{\rho}_{0}=\rho(X_{1},X_{2}|S_{0}^{n}Q)$,
$\rho_{12|0}=\rho(S_{1},S_{2}|S_{0})=\frac{\rho_{12}-\rho_{01}\rho_{02}}{\sqrt{\left(1-\rho_{01}^{2}\right)\left(1-\rho_{02}^{2}\right)}}$
and $\theta_{k}^{\prime}=\theta\left(X_{k},U^{n}|S_{0}^{n}Q\right),k=1,2$.
Then utilizing Lemmas \ref{lem:iid}, \ref{lem:Gaussian} and \ref{lem:dpi},
we have
\begin{equation}
\hat{\rho}_{0}\leq\rho_{m}(S_{1}^{n},S_{2}^{n}|S_{0}^{n}Q)=\rho_{m}(S_{1},S_{2}|S_{0})=\rho_{12|0},
\end{equation}
\begin{align}
\hat{\rho}_{0} & \leq\theta\left(X_{1},S_{2}^{n}|S_{0}^{n}Q\right)\\
 & \leq\theta\left(X_{1},U^{n}|S_{0}^{n}Q\right)\rho_{m}\left(S_{2}^{n},U^{n}|S_{0}^{n}Q\right)\\
 & =\theta_{1}^{\prime}\beta_{2},\label{eq:-4}
\end{align}
and
\begin{equation}
\hat{\rho}_{0}\leq\theta_{2}^{\prime}\beta_{1}.
\end{equation}

Now based on the inequalities above and utilizing the outer bound
$\mathcal{R}_{2}^{(o)}$ of Theorem \ref{thm:AdmissibleRegion-MAC},
we can obtain a sequence of desired results. Specifically, Combining
the inequality $I(S_{1}S_{2};\hat{S}_{1}\hat{S}_{2})\leq I\left(X_{1}X_{2};Y|Q\right)$
in $\mathcal{R}_{2}^{(o)}$ with
\begin{equation}
I(S_{1}S_{2};\hat{S}_{1}\hat{S}_{2})\geq R_{S_{1}S_{2}}(D_{1},D_{2})
\end{equation}
and
\begin{align}
I\left(X_{1}X_{2};Y|Q\right) & =h\left(Y|Q\right)-h\left(Y|X_{1}X_{2}\right)\\
 & \leq h\left(Y\right)-h\left(Y|X_{1}X_{2}\right)\\
 & \leq\frac{1}{2}\log\left(1+\textrm{var}(X_{1}+X_{2})\right)\\
 & =\frac{1}{2}\log\bigl(1+\textrm{var}(X_{1})+\textrm{var}(X_{2})\nonumber \\
 & \qquad+2\rho(X_{1},X_{2})\sqrt{\textrm{var}(X_{1})\textrm{var}(X_{2})}\bigr)\\
 & \leq\frac{1}{2}\log\left(1+P_{1}+P_{2}+2\hat{\rho}\sqrt{P_{1}P_{2}}\right),
\end{align}
gives
\begin{equation}
R_{S_{1}S_{2}}(D_{1},D_{2})\leq\frac{1}{2}\log\left(1+P_{1}+P_{2}+2\hat{\rho}\sqrt{P_{1}P_{2}}\right).
\end{equation}

In addition, from Property \eqref{eq:i} of Lemma \ref{lem:properties},
we have
\begin{align}
1-\theta^{2}\left(X_{1},S_{0}^{n}Q\right) & =\frac{1-\theta^{2}\left(X_{1},X_{2}S_{0}^{n}Q\right)}{1-\theta^{2}\left(X_{1},X_{2}|S_{0}^{n}Q\right)}\\
 & \leq\min\left(\frac{1-\hat{\rho}^{2}}{1-\rho_{12|0}^{2}},1\right),\label{eq:-6}
\end{align}
where the inequality \eqref{eq:-6} follows from
\begin{equation}
\theta\left(X_{1},X_{2}S_{0}^{n}Q\right)\geq\theta\left(X_{1},X_{2}\right)\geq\rho\left(X_{1},X_{2}\right),
\end{equation}
and
\begin{equation}
\theta\left(X_{1},X_{2}|S_{0}^{n}Q\right)\leq\rho_{m}\left(X_{1},X_{2}|S_{0}^{n}Q\right)\leq\rho_{12|0}.
\end{equation}
Then combining the inequality $I(S_{1}S_{2};\hat{S}_{1}\hat{S}_{2}|S_{0})\leq I\left(X_{1}X_{2};Y|S_{0}^{n}Q\right)$
in $\mathcal{R}_{2}^{(o)}$ with
\begin{equation}
I(S_{1}S_{2};\hat{S}_{1}\hat{S}_{2}|S_{0})\geq R_{S_{1}S_{2}|S_{0}}(D_{1},D_{2})
\end{equation}
and{\small{}{}
\begin{align}
 & I\left(X_{1}X_{2};Y|S_{0}^{n}Q\right)\nonumber \\
 & =h\left(Y|S_{0}^{n}Q\right)-h\left(Y|X_{1}X_{2}\right)\nonumber \\
 & \leq\frac{1}{2}\log\left(1+\mathbb{E}\textrm{var}(X_{1}+X_{2}|S_{0}^{n}Q)\right)\nonumber \\
 & =\frac{1}{2}\log\Bigl(1+\mathbb{E}\textrm{var}(X_{1}|S_{0}^{n}Q)+\mathbb{E}\textrm{var}(X_{2}|S_{0}^{n}Q)\nonumber \\
 & \qquad+2\rho(X_{1},X_{2}|S_{0}^{n}Q)\sqrt{\mathbb{E}\textrm{var}(X_{1}|S_{0}^{n}Q)\mathbb{E}\textrm{var}(X_{2}|S_{0}^{n}Q)}\Bigr)\nonumber \\
 & \leq\frac{1}{2}\log\Bigl(1+\left(1-\theta^{2}\left(X_{1},S_{0}^{n}Q\right)\right)P_{1}+\left(1-\theta^{2}\left(X_{2},S_{0}^{n}Q\right)\right)P_{2}\nonumber \\
 & \qquad+2\hat{\rho}_{0}\sqrt{\left(1-\theta^{2}\left(X_{1},S_{0}^{n}Q\right)\right)\left(1-\theta^{2}\left(X_{2},S_{0}^{n}Q\right)\right)P_{1}P_{2}}\Bigr)\nonumber \\
 & \leq\frac{1}{2}\log\biggl(1+\min\Bigl(\frac{1-\hat{\rho}^{2}}{1-\rho_{12|0}^{2}},1\Bigr)\left(P_{1}+P_{2}+2\hat{\rho}_{0}\sqrt{P_{1}P_{2}}\right)\biggr)
\end{align}
}gives{\small{}{}
\begin{align}
 & R_{S_{1}S_{2}|S_{0}}(D_{1},D_{2})\nonumber \\
 & \leq\frac{1}{2}\log\biggl(1+\min\Bigl(\frac{1-\hat{\rho}^{2}}{1-\rho_{12|0}^{2}},1\Bigr)\left(P_{1}+P_{2}+2\hat{\rho}_{0}\sqrt{P_{1}P_{2}}\right)\biggr).
\end{align}
}Similarly, the last five inequalities in \eqref{eq:outerboundGaussian}
can be obtained as well. This completes the proof.

\end{document}